\documentclass[a4paper,10pt,twocolumn,twoside]{article}

\usepackage{graphicx,natbib,aas_macros}
\usepackage{fancyhdr}

\tt

\title{Novae: I. The maximum magnitude relation with decline time (MMRD) and distance}

\author{Nimisha G. Kantharia \\
\normalsize National Centre for Radio Astrophysics, \\ 
\normalsize Tata Institute of Fundamental
Research, \\ \normalsize Post Bag 3, Ganeshkhind, Pune-411007, India   \\
\normalsize \it Email: nkprasadnetra@gmail.com \\
\normalsize \it URL: https://sites.google.com/view/ngkresearch/home}

\date{March 2017}

\begin{document}
\maketitle

\thispagestyle{empty}

\begin{abstract}
{\small The origin and calibration of the maximum absolute magnitude relation with decline time 
(MMRD) for novae, first derived by \citet{1936PASP...48..191Z}, empirically validated by 
\citet{1940ApJ....91..369M} and widely used to estimate distances to 
classical novae and the near-constancy of the absolute magnitude of novae, 15 days after optical
maximum, suggested by \citet{1955Obs....75..170B} are revisited in this paper and found to 
be valid.  The main results presented in the paper
are:  (1) A physical derivation of the MMRD based on instantaneous injection of energy
to the nova system. (2) A significantly better-constrained MMRD: 
$\rm \bf M_{V,0} = 2.16(\pm0.16)log_{10}t_2 - 10.804(\pm0.117)$ using a 
two step calibration procedure.  (3) It is shown that the MMRD is one
of the best distance estimators to novae available to us and that accuracy of the
distances is predominantly limited by an underestimated peak apparent brightness.
(4) It is shown that the same MMRD calibration is applicable to novae of all speed class and to both
Galactic and extragalactic novae. (5) It is shown that the absolute magnitudes of novae 
with $\rm 2.4 \le t_2 \le 86$ days have a smaller scatter on day 12 
($\rm \bf M_{V,12} = -6.616 \pm 0.043$) compared to day 15 following optical maximum. 

We reiterate the need for homogenised high fidelity spectrophotometric data in optical bands 
on classical and recurrent novae in outburst 
to effectively utilise the potential of the MMRD and $\rm M_{V,12}$ in determining their 
luminosities and distances.  }
\end{abstract}

\section*{\small Keywords}
Classical novae, recurrent novae, MMRD,  absolute magnitude, light curves,  
emission line widths, distance to novae. 

\section{Introduction}
It is now well-established that
novae are stellar binaries consisting of a white dwarf known as the primary and
a gaseous companion star known as the secondary which is either a main sequence star, 
a sub-giant or
a red giant star.  Novae are classified into three types - classical novae for which only 
one major outburst has been recorded, recurrent novae for which more than one major outburst
has been recorded and dwarf novae which show low amplitude frequent outbursts.
During an outburst in classical and recurrent novae, the quiescent system can brighten by
$8-20$ magnitudes in a day or so whereas a dwarf nova system brightens by $2-6$ magnitudes
in a short time.  The physical mechanism responsible for the brightening of dwarf novae 
is believed to be different from that for classical and recurrent novae.  While
classical and recurrent novae show evidence for ejection of matter by the white dwarf,
no such evidence has been found for dwarf novae.  

Estimating distances to astronomical objects has always been a difficult task.  
Distances to novae are also difficult to estimate and suffer from large uncertainties
which in turn lead to uncertainties in the intrinsic luminosity of novae outburst
when estimated by using the distance.  
Several methods have been employed to estimate distances to classical novae and these have been
explained in 
\citet{1942PA.....50..233M}.  Expansion parallax, intensities of Calcium lines,
Galactic rotation contribution to the velocities of interstellar Calcium lines, 
reddening with distance etc are some methods which have been used to estimate distances to novae. 
A reliable distance estimator which is often used is the observed angular expansion of ejecta
shells around old classical novae alongwith their expansion velocities although this 
can only be used for novae for which the explosion has left behind a detectable shell. 
Distances are also estimated 
using the maximum (absolute) magnitude relation with decline time (MMRD)
and it has been remarkably successful as a statistical tool since the uncertainties in
the peak absolute magnitude and distance estimate for a given nova are found to be large.  
In this method, the peak luminosity of the nova outburst can be estimated without knowing
the distance to the nova. The distance is then determined using the 
maximum luminosity and the observed peak apparent brightness of the nova. 
However the MMRD needs to be calibrated before it can be used
to estimate the peak absolute magnitudes of novae. 

That a relation 
between decline time and maximum luminosity should exist for novae outbursts was first suggested by
\citet{1936PASP...48..191Z} under the assumption that
there is an instantaneous release of energy in a nova explosion. 
Zwicky found that the peak visual luminosity can be inferred from both the
ejecta velocity ie expansion-luminosity
relation \citep{1936PNAS...22..457Z} and the decline time ie life-luminosity relation
\citep{1936PASP...48..191Z} which he then combined to get a life-expansion relation.
However there were errors in these relations which were pointed out 
and corrected by \citet{1940ApJ....91..369M} who also presented 
empirical evidence in support of the corrected relation. 
In the corrected MMRD, the most luminous nova outbursts are the fastest ones
i.e. novae in which the decline in luminosity is most rapid. 
Since then the relation has been widely calibrated 
\citep[e.g.][]{1956AJ.....61...15A,1957ZA.....41..182S,1960stat.conf..585M,
1985ApJ...292...90C, 1995ApJ...452..704D,2000AJ....120.2007D} and used as a distance estimator.
While the importance and potential of the MMRD for estimating distances has been
widely appreciated, its physical origin continues to baffle us.
To understand and possibly rectify this, we explored a physical treatment based on
the original assumption of rapid injection of energy which we find works
remarkably well in deriving the life-expansion relation and the MMRD. 
A two-step calibration procedure using available observational data on novae results in
a more reliable MMRD.   Novae are classified into two main speed classes namely 
fast (or flashing) and slow based on their rate of decline \citep{1936PA.....44...78G}.
Since decline rates of novae of different 
speed classes are not same, \citet{1955Obs....75..170B} explored the possibility
that their light curves would intersect on some day following the optical maximum
and presented evidence for the same.  With more data now available, we revisit this
in the paper.  We use the data on classical, recurrent and dwarf novae in our Galaxy
from the following papers in literature:
\citet{1960stat.conf..585M, 1983ApJ...268..689C, 1985ApJ...292...90C, 1987MNRAS.227...23W, 
1995warner.book.....W, 2000AJ....120.2007D, 2010ApJS..187..275S, 2010AJ....140...34S, 2011ApJS..197...31S}. 
Data on extragalactic novae are used from 
\citet{1956AJ.....61...15A, 1964AnAp...27..498R, 1987ApJ...318..507P, 2016ApJS..227....1S}.

\section{The MMRD}
In this section, we discuss the physical origin of the MMRD for classical (and
recurrent) novae,  devise a better method to 
calibrate it based on Zwicky's suggestion and then use the relation 
to estimate the peak absolute magnitude of several novae.

\subsection{The origins of the MMRD}
The MMRD, originally referred to as the life-luminosity relation was first suggested and
derived by \citet{1936PNAS...22..457Z,1936PASP...48..191Z} under the assumption of
instantaneous release of energy in a nova outburst.  `Life' here refers to the decline time. 
He derived the life-luminosity (peak absolute magnitude $\rm M_{0}$) relation as:
\begin{equation}
\rm
M_{0}=-5 log(t_{\Delta m}) + constant
\label{eqn1}
\end{equation}
where $\rm t_{\Delta m}$ is defined as the time in which the nova is brighter
than $\rm M_{0}+\Delta m$.  Based on the observational data available at that time,
Zwicky suggested that $\rm \Delta m$ should be taken to
be 2 or 3.  Hence $\rm \Delta m=1$ should also work if data quality and temporal sampling
are sufficient. 
He also obtained the expansion-luminosity relation ie
between the  ejecta velocity $\rm \Delta v$ and $\rm M_{0}$ as:
\begin{equation}
\rm
M_{0}=-5 log(\Delta v) + constant
\label{eqn2}
\end{equation}
The above two expressions were combined to obtain a relation between observables 
\citep{1936PASP...48..191Z} ie the life-expansion relation:
\begin{equation}
\rm
log(t_{\Delta m}) = log(\Delta v) + constant
\label{eqn3}
\end{equation}
\citet{1936PASP...48..191Z} calibrated this relationship using some of the available 
data in 1936 and then
transferred it to Equation \ref{eqn1} to estimate the peak absolute magnitude and
noted that this relation would be useful in determining the distance to a nova.
However he combined data on novae and supernovae which resulted in an error in
calibration of the relation.  Thus, Zwicky deduced that slow novae are intrinsically
more luminous than fast novae which was contrary to results from observations of novae.  
This calibration error was explained and resolved by
\citet{1940ApJ....91..369M} who also provided a calibration of the
relation in Equation \ref{eqn3} from observational data on novae. 
Thus, while Zwicky found a direct relation between the $\rm t_{\Delta m}$ and $\rm \Delta v$,
observations were explained by an inverse relation \citep{1940ApJ....91..369M}. 
Using the velocity displacements $\rm V$ of the principal series of
absorption lines as representative of the ejecta velocity, \citet{1940ApJ....91..369M} found that 
\begin{equation}
\rm log V = 3.19 - 0.49~ log~ t 
\label{eqn4}
\end{equation}
Moreover from observational data on novae, \citet{1940ApJ....91..369M} arrived
at three important conclusions on spectral and light curve evolution of novae 
which, like the MMRD, have stood the test of time and been 
borne out by subsequent observations.  These were: (1) that the observed spectral stage of a
nova depends on the light curve evolution; (2) that the decline time is
inversely proportional to the square of the velocity of the principal spectrum (Equation \ref{eqn4}) and (3) that
velocity of the diffuse enhanced spectrum is generally twice the principal spectrum velocities.
\citet{1960stat.conf..585M} and \citet{1957gano.book.....G} 
presented extensive summaries and interpretation of the photometric and spectroscopic data 
in the optical bands on novae. 

\subsection{The physical origin of the MMRD}
\label{sec-origin}
Here we revisit the origin of the MMRD using simple physical arguments.  We start
with the original assumption under which \citet{1936PASP...48..191Z} derived the
life-luminosity relation (Equation \ref{eqn1}) namely that most of the energy in a nova outburst 
is released in a short time.  This assumption is validated by the extremely rapid rise 
($\sim$ day) in the optical emission
to maximum ($\sim 10^3-10^8$ times increase in quiescence luminosity) detected in  
classical and recurrent novae.  We then derive the life-expansion and life-luminosity relations.
We, especially, note that the relation between the
expansion velocity and decline time that was empirically found by 
\citet{1940ApJ....91..369M} naturally follows from the physical arguments presented here.  
We proceed as follows.   

The total energy $\rm E_{tot}$ released in the cataclysmic thermonuclear 
explosion on the white dwarf is immediately released in the form of kinetic energy imparted to the 
matter overlying it in the relatively lower density envelope on the surface of
the degenerate white dwarf.  This, in a general case,
can lead to two physical manifestations - a macroscopic bulk motion which 
accelerates and ejects 
matter and a microscopic random motion component which leads to localized
motions of the constituent atoms and molecules in the ejecta.
Due to the first component, the ejecta starts to expand and we refer to this energy component
as $\rm E_{kin}$.  Due to the second component, the ejecta material can get 
collisionally ionized and start radiating.  
We refer to this radiating energy component as $\rm E_{rad}$.  Both 
are a result of the kinetic energy imparted to the ejecta by the explosion.
So larger is $\rm E_{tot}$, larger are $\rm E_{kin}$ and $\rm E_{rad}$ although they need
not be in equipartition.   Thus, just after the explosion: 
\begin{equation}
\rm
E_{tot} = E_{kin} + E_{rad}; ~~~ E_{kin} \propto E_{rad}
\label{eqn5}
\end{equation}
In rest of this discussion, we are concerned with the parameters near peak optical emission
when the optical depth of the ejecta becomes one as it expands, soon after the outburst.  
Thus the quantities like luminosity and ejecta velocity used below refer to values at or
close to the peak optical emission.  More energetic the outburst,
faster will be the expansion of the ejecta.
The maximum kinetic power will be proportional to the kinetic energy imparted to the nova 
ejecta which leads to its expansion.  
Thus,
\begin{equation}
\rm
L_{kin} \propto E_{kin} = \frac{1}{2} m_{ej} v_{ej}^2;~~~  L_{kin} \propto v_{ej}^2
\label{eqn6}
\end{equation}
where $\rm v_{ej}$ is the expansion velocity of the ejecta of mass $\rm m_{ej}$.
For now we assume that there is no systematic effect of variation in
$\rm m_{ej}$ on $\rm L_{kin}$ although we are aware that the range of $\rm m_{ej}$ can
lead to a scatter on the MMRD.  This and several other arguments are
motivated by the apriori information that the MMRD has been supported by
observational data on novae.  We, thus, probe 
the physics behind MMRD with appropriate assumptions where required.    

The maximum radiative energy of the nova will determine its radiative luminosity
which is often estimated to be close to the Eddington luminosity at the peak of the 
optical light curve.  Given $\rm E_{rad}$,
we can estimate $\rm L_{rad}$ by dividing the former by the time during which the
nova radiates close to its peak luminosity.  Since we do not know this time,
another way to get a lower limit to $\rm L_{rad}$ is to divide the energy by $\rm t_2$ which
is the time taken by the light curve to fall by 2 magnitudes from the peak.
Another way to understand this is that the average luminosity of the nova 
when it is brighter than $\rm M_{V,0}+2$ can be estimated by
dividing the maximum radiated energy by $\rm t_2$.   This gives a lower limit to the
peak luminosity.  Thus if $\rm E_{rad}$ is a constant, then this means that
$\rm L_{rad}$ will be larger for faster novae.  
\begin{equation}
\rm
L_{rad} > \frac{E_{rad}}{t_2};~~~  L_{rad} \propto t_2^{-1}
\label{eqn7}
\end{equation}
For practical purposes, we use the proportionality relation.
For now, we assume no systematic relation between $\rm E_{rad}$ and $\rm L_{rad}$ exists
and also note that if the derivation and calibration of MMRD presented here succeed in explaining
observations, then it would give strong support to Equation \ref{eqn5}.  Equation \ref{eqn7}
then results in the life-luminosity relation or the MMRD since 
the peak absolute magnitude is:
\begin{equation}
\rm 
M_0 = -2.5 log_{10} L_{rad}  + K~~or~
M_{0} = - 2.5~log_{10} t_2^{-1} + K_1 
\label{eqn8}
\end{equation}
This is different from Equation \ref{eqn1} which Zwicky (1936) had derived. 
From Equations \ref{eqn5}, \ref{eqn6}, \ref{eqn7}, for the rapid induction of energy 
in the nova system, we have:
\begin{equation}
\rm
L_{kin} \propto L_{rad};i.e.~~~v_{ej}^2 \propto t_{2}^{-1}
\label{eqn9}
\end{equation}
This is the relation that \citet{1940ApJ....91..369M} had deduced 
(see Equation \ref{eqn4}) from
observations of the principal spectrum lines and the light curve as mentioned in the previous 
subsection.  Equation \ref{eqn9} gives the life-expansion relation i.e.
\begin{equation}
\rm
log_{10} t_2^{-1} =  2~log_{10}(v_{ej}) + A_2
\label{eqn10}
\end{equation}
This equation is different from what \citet{1936PASP...48..191Z} had derived as given in 
Equation \ref{eqn3}.  Since larger volumes of data are now available to determine the constants in the
expression, we can write the life-expansion relation in a more general form:
\begin{equation}
\rm
log_{10} t_2 =  A_1~log_{10}(v_{ej}) + A_2
\label{eqn11}
\end{equation}
where $\rm v_{ej}$ can be determined from the widths of emission lines recorded near optical peak. 
From Equation \ref{eqn9}, we can write $\rm L_{rad} \propto v_{ej}^2$ 
which then gives us the expansion-luminosity relation as follows:
\begin{equation}
\rm M_{0} = - 2.5~log_{10} v_{ej}^2 + K_2 = -5~log_{10}v_{ej} + K_2 
\label{eqn12}
\end{equation}
This equation is similar to Equation \ref{eqn2} that \citet{1936PASP...48..191Z} had derived. 
The match with observations, as we demonstrate in the next sections, suggest that 
the physical treatment of 
the energetics of a nova explosion presented here and the assumptions implicit in 
deriving these encapsulate the macroscopic properties of a nova explosion.    

\subsection{Calibrating the MMRD}

We use the life-luminosity and life-expansion relations of
Equations \ref{eqn8} and \ref{eqn11} to calibrate the MMRD.
We determine the constants $\rm K_1, A_1, A_2$ from observational data in literature 
following a two-step procedure. 
We start with the relation in Equation \ref{eqn11} and use observational data
on emission line widths ($\rm \Delta v$) 
estimated near the optical peak as indicative of the ejecta velocity and $\rm t_2$ 
to determine the constants $\rm A_1, A_2$.  
The data on $\rm \Delta v$ and $\rm t_2$ for 51 novae are taken from \citet{2011ApJS..197...31S} and
these are plotted in Figure \ref{fig1} and listed in Table \ref{tab2} in the Appendix.  
The  $\rm \Delta v$ have been estimated from the widths of
the H$\alpha$ or H$\beta$ emission lines near visible maximum \citep{2011ApJS..197...31S}.
$\rm t_2$ is in days and $\rm \Delta v$ is in kms$\rm ^{-1}$.  The best least squares fit to the data
is obtained when $\rm A_1 = -0.8655\pm0.15$ and $\rm A_2 = 3.8276\pm0.51$ which is
shown by the line in Figure \ref{fig1}.
Thus, the life-expansion relation which encapsulates their inverse relation is:
\begin{equation}
\rm \bf
log_{10}~t_2 = -0.8655(\pm 0.15) log_{10}~\Delta v + 3.8276(\pm 0.51)
\label{eqn13}
\end{equation}
We keep in mind that the signs of the coefficients will be the same when transferred to
the $\rm t_2 \rightarrow L_{rad}$ relation and will change when the MMRD in magnitudes is estimated. 
The 51 data points used to derive the above relation also
includes three novae in the Large
Magellanic Cloud (LMC), the very slow nova V723 Cas and recurrent novae.  

For the same data shown in Figure \ref{fig1}, we can fit the functional form: 
$\rm log_{10} \Delta v = -0.44 (\pm0.08)~log_{10} t_2 + 3.7(\pm0.09) $  which compares
remarkably well with the relation $\rm log_{10} V = - 0.5~log_{10} t_2 + 3.57$  
derived by \citet{1960stat.conf..585M} using the line shifts of the principal
absorption lines.
$\rm t_2$ is in days and $\rm \Delta v$, V are in kms$\rm ^{-1}$.  These relations are not
very different from Equation \ref{eqn4} obtained in 1940 giving strong support to the correlation
which has stood the test of time and has been verified on larger volumes of data.
We also fitted the available data on $\rm \Delta v$ and $\rm t_3$ 
for 22 novae listed in \citet{2010AJ....140...34S}.  The best least square fit
gives the relation $\rm log_{10} t_3 = -1.08 (\pm0.28) log_{10} \Delta v + 5.01 (\pm 0.95)$.
We continue the calibration of the MMRD with Equation \ref{eqn13}. 

\begin{figure}
\begin{center}
\includegraphics[width=9.0cm]{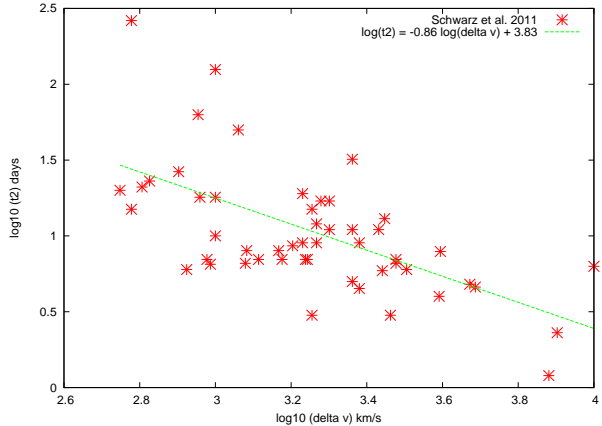}
\caption{\small The observed line width $\rm \Delta v$ is plotted against the time
taken by a nova to decline in brightness by 2 magnitudes $\rm t_2$.  The data have
been taken from \citet{2011ApJS..197...31S}.  The line shows the best fit to
the data which is $\rm log_{10}t_2 = -0.86 log_{10} \Delta v + 3.83$.  This is
step one in calibrating the MMRD. }
\label{fig1}
\end{center}
\end{figure}

At the end of the first step, we have calibrated the relation between the ejecta velocity
and decline time.  In the second step, we transfer this to the MMRD in Equation \ref{eqn8}
and fine-tune the zero-point i.e. estimate $\rm K_1$.  
Substituting from Equation \ref{eqn13} and retaining $\rm log_{10} t_2$ as the variable, we have
$\rm M_{V,0} = -2.5~(-0.8655~log_{10}t_2 + 3.8276) + K_1$ i.e.:
\begin{equation}
\rm 
M_{V,0} = 2.16(\pm 0.15)~log_{10}t_2 - 9.57(\pm 0.51) + K_1.
\label{eqn14}
\end{equation}
This relation with $\rm K_1=0$ alongwith the one sigma uncertainties are plotted with lines in 
Figure \ref{fig2}(a).  The overplotted $\sim 60$ points show the peak absolute magnitude of
several novae in the V band estimated using other methods.  These data have been taken
from \citet{ 1983ApJ...268..689C, 1985ApJ...292...90C, 2000AJ....120.2007D,
2010ApJS..187..275S}.   There is clearly an offset between
the peak absolute magnitudes predicted by the fit as shown by the lines and the data 
since we have set $\rm K_1=0$.  To estimate $\rm K_1$, we used the 
same data plotted in Figure \ref{fig2}(a) since these have been carefully estimated 
in most cases for calibrating the MMRD.
We then fixed the rate of change in $\rm M_{V,0}$ with $\rm log_{10}t_2$ 
i.e. the coefficient of $\rm log_{10}t_2$ to 2.16 as given in Equation \ref{eqn14} and
started with the initial guess of -9.57 for the intercept.  The best least squares fit is
is the following and which is shown in Figure \ref{fig2}(b) along with $1 \sigma$ errors:
\begin{figure*}
\centering
\includegraphics[width=10.5cm]{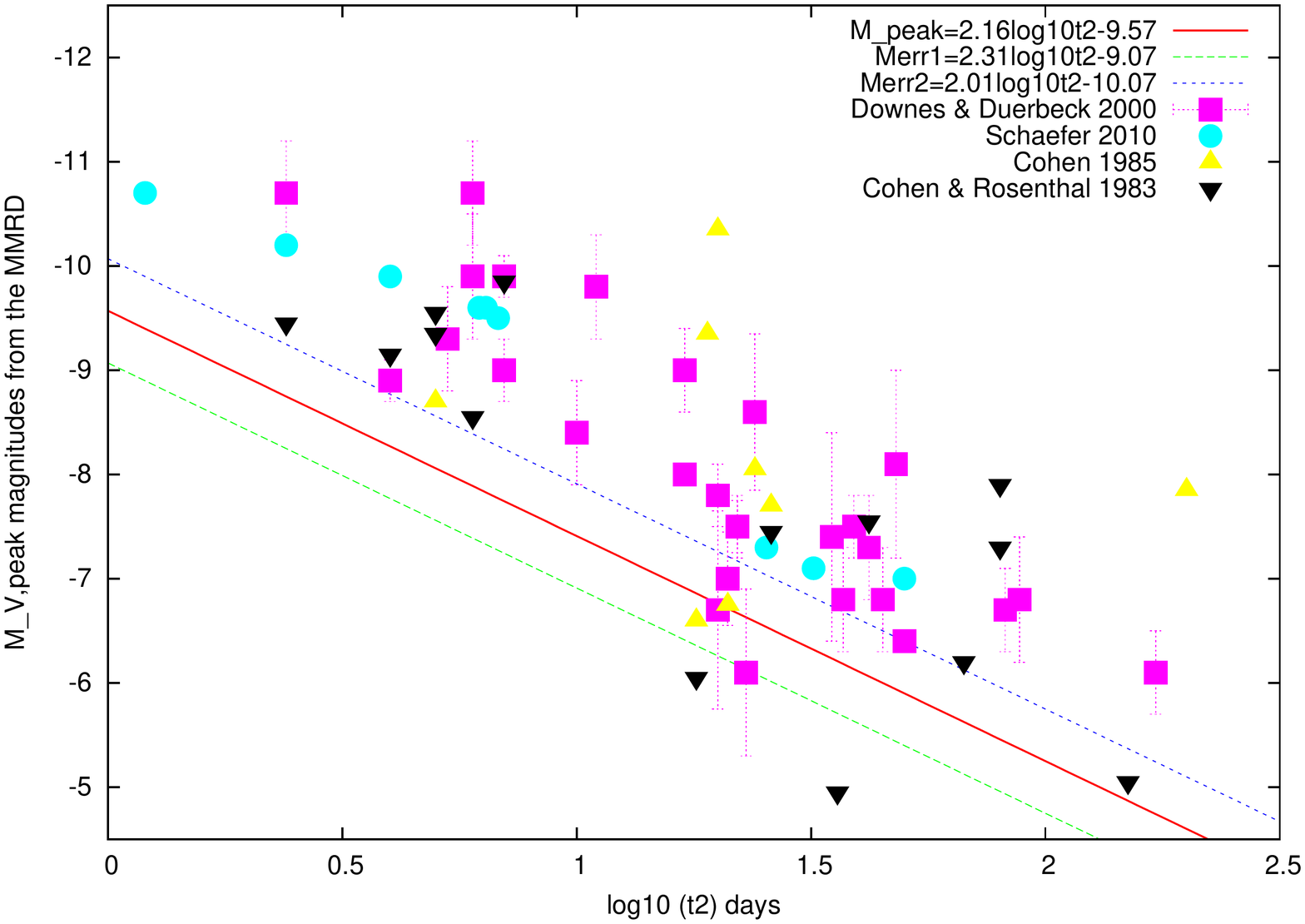}(a)
\includegraphics[width=10.5cm]{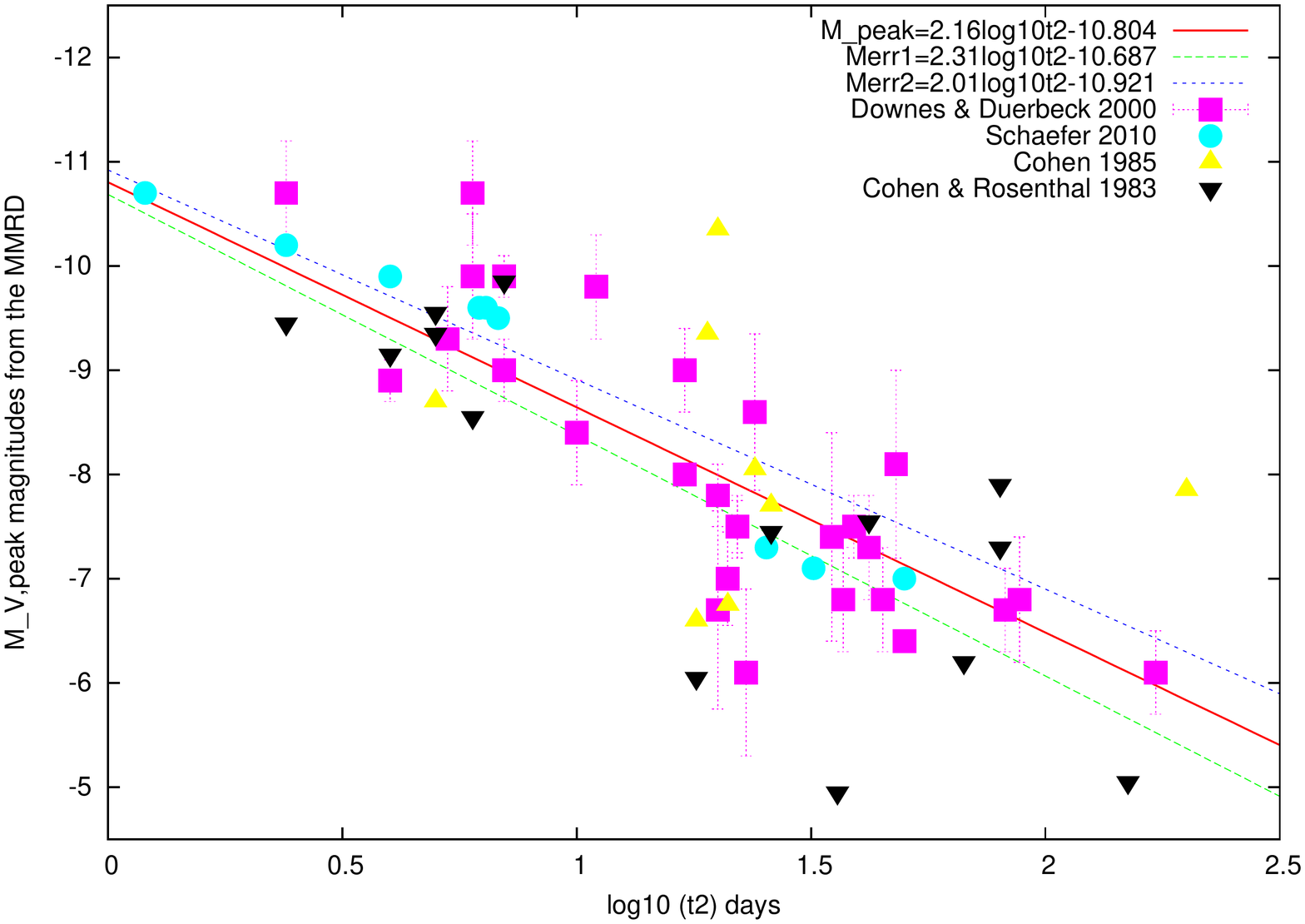}(b)
\caption{\small Step two in calibrating the MMRD.  The data on the peak absolute magnitudes
on about 60 novae which have been derived using mostly methods other than MMRD are plotted 
with a distinct symbol corresponding to one of the references - 
\citet{1983ApJ...268..689C, 1985ApJ...292...90C, 2000AJ....120.2007D, 2010ApJS..187..275S}. 
(a) The MMRD, after the calibration shown in Figure \ref{fig1} is transferred via Equation \ref{eqn8}, is
shown by the solid line.  The dashed lines represent the $1 \sigma$ errors on the
MMRD peak absolute magnitudes. 
Note the offset between the MMRD-determined peak absolute magnitdues and the data.
(b) The final calibrated MMRD after determining the correct zero-point.
$\rm M_{V,peak}=2.16log_{10}t_2 - 10.804$ and the dashed lines on either side are the $1 \sigma$ errors.  
Note the lower uncertainties on this MMRD compared to (a) and the offset between 
the data and the fit has now been calibrated.} 
\label{fig2}
\end{figure*}
\begin{equation}
\rm \bf
M_{V,0} = 2.16(\pm 0.15)~log_{10}t_2 - 10.804(\pm 0.117).  
\label{eqn15}
\end{equation}
The fit from Equation \ref{eqn15} in Figure \ref{fig2}(b) is remarkably good
and the scatter on the determination of the absolute magnitude is less than $0.5$ 
magnitudes especially for fast novae. 
That the MMRD presented here is the best calibrated and reliable relation is demonstrated 
by comparing with the MMRD calibrations that exist in literature: 
\begin{enumerate}
\item $\rm M_{V,0} = -11.5 + 2.5~log~t_3$ \\
by \citet{1957ZA.....41..182S,1960stat.conf..585M}
\item  $\rm M_{V,0} = -10.66(\pm0.33) + 2.31(\pm0.26)~logt_2$ \\
by \citet{1985ApJ...292...90C}.
\item $\rm M_{V,0} = -7.92 - 0.81~arctan \frac{1.32-log~t_2}{0.23}$ \\
by \citet{1995ApJ...452..704D}.
\item $\rm M_{V,0} = -11.32 (\pm 0.44) + 2.55 (\pm 0.32) log~t_2$ \\ and
$\rm M_{V,0} = -8.02 - 1.23~arctan \frac{1.32-log~t_2}{0.23}$ \\
by \citet{2000AJ....120.2007D}. 
\end{enumerate}
Clearly the MMRD calibrated using the two-step procedure has resulted
in a well-constrained MMRD.   The first step exclusively uses directly observed 
quantities.   The calibration has also used larger datasets than previously employed.
Figure \ref{fig2} demonstrates that a carefully calibrated MMRD can result in well-constrained 
peak absolute magnitude of a nova and that a single MMRD can be used for all speed classes
of classical and recurrent novae.  It has helped that we were aware of the success of
the MMRD in determining the peak absolute magnitudes of novae in outburst which was
not the case in the early days of the MMRD.  
\begin{figure}[h]
\centering
\includegraphics[width=8.0cm]{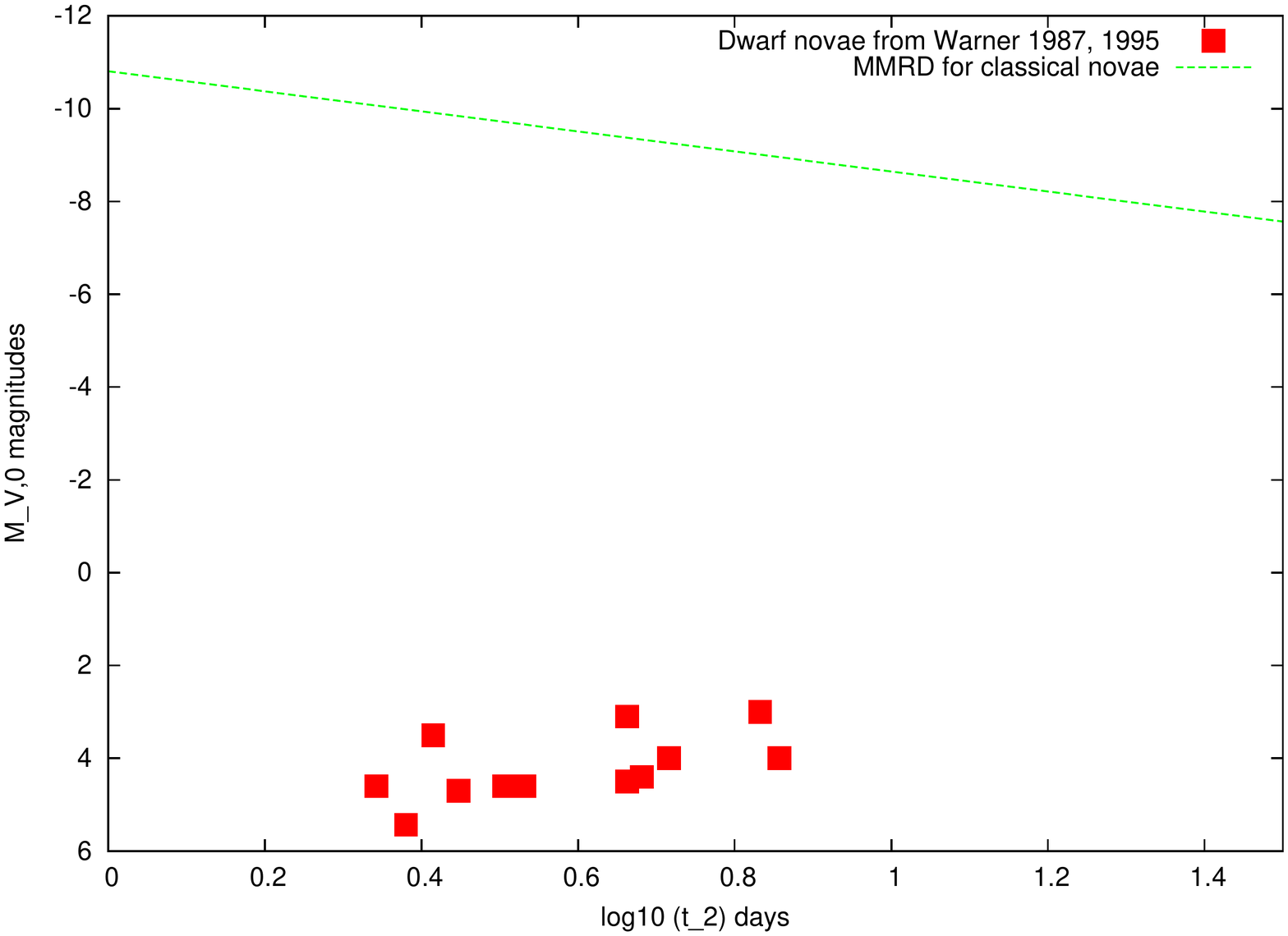}(a)
\includegraphics[width=8.0cm]{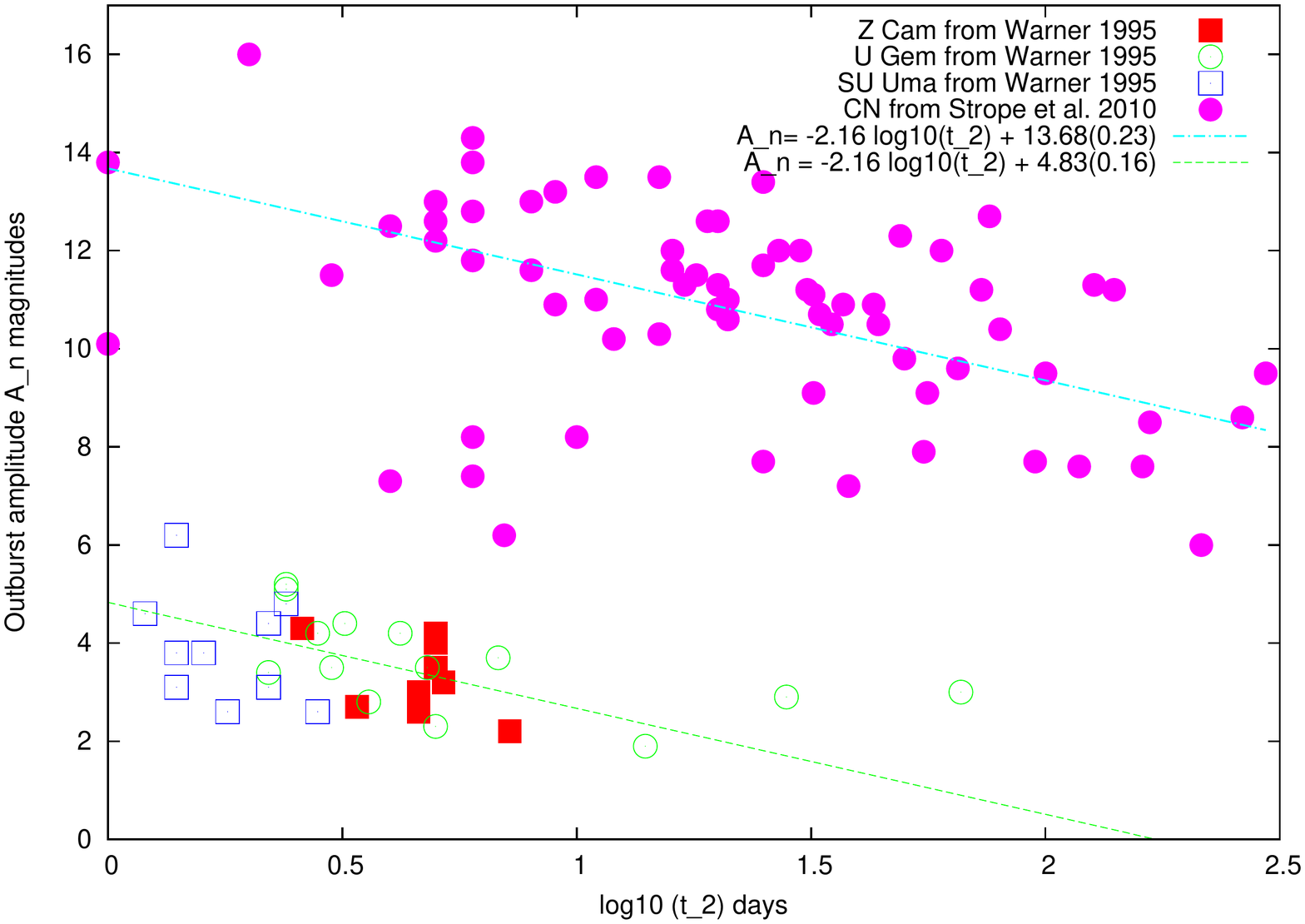}(b)
\caption{\small (a) The MMRD of Equation \ref{eqn15} for classical novae is shown as the solid
line and the data on dwarf novae taken from \citet{1987MNRAS.227...23W,1995warner.book.....W}
are shown by the symbols.  Dwarf novae do not follow the MMRD.
(b) Figure shows the outburst amplitude of classical novae (filled circles) \citep{2010AJ....140...34S}  
and for the three types of dwarf novae \citep{1995warner.book.....W}.  Overplotted on each
dataset is the MMRD-like relation with the same rate of change in $\rm M_{V,0}$ with $\rm t_2$.  
The zero-points have been independently estimated for the two types of novae.  Note the
good match between data and MMRD-like relation for both types of novae.  }
\label{fig3}
\end{figure}
\begin{figure}[t]
\centering
\includegraphics[width=8.0cm]{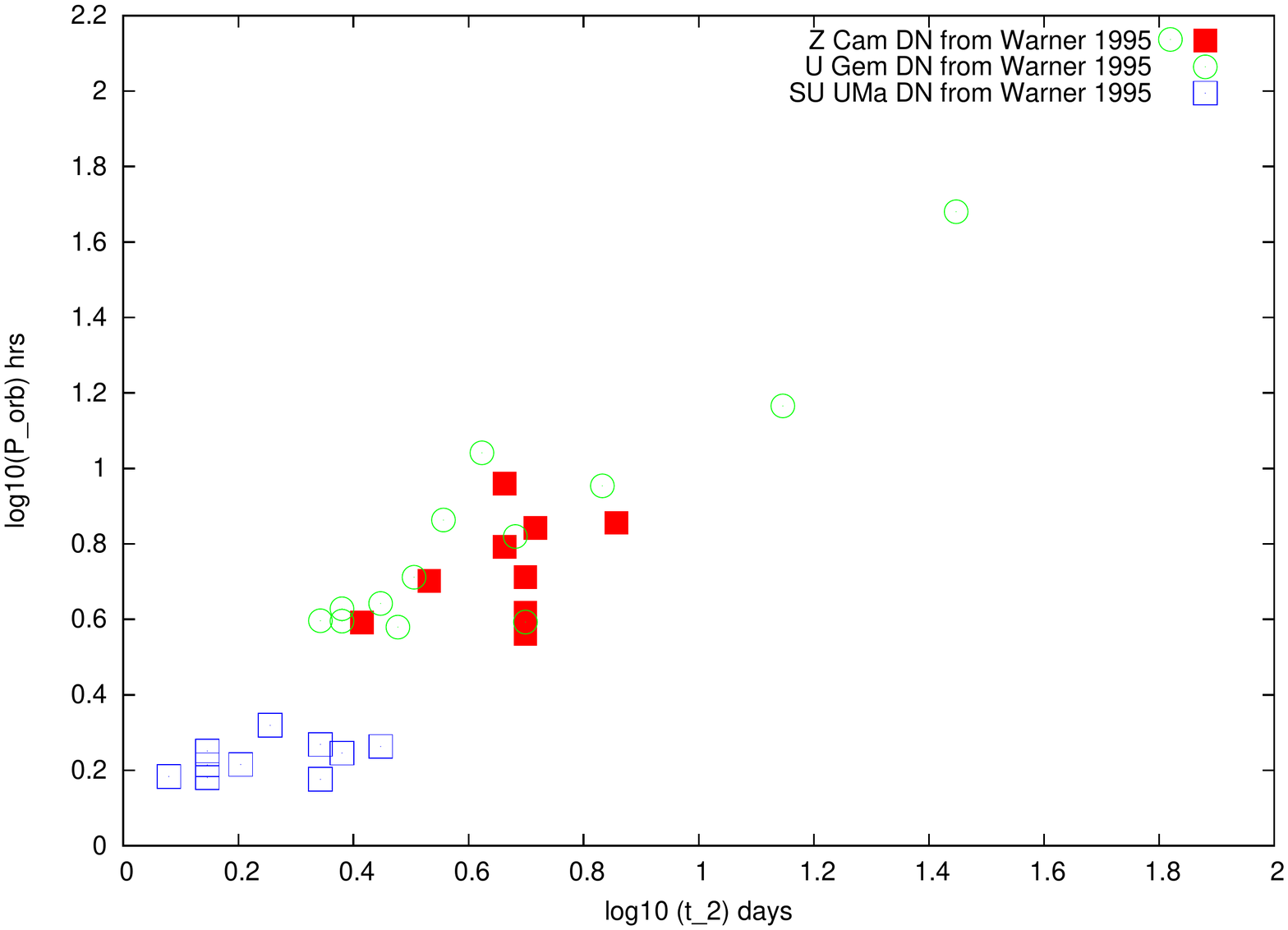}(a)
\includegraphics[width=8.0cm]{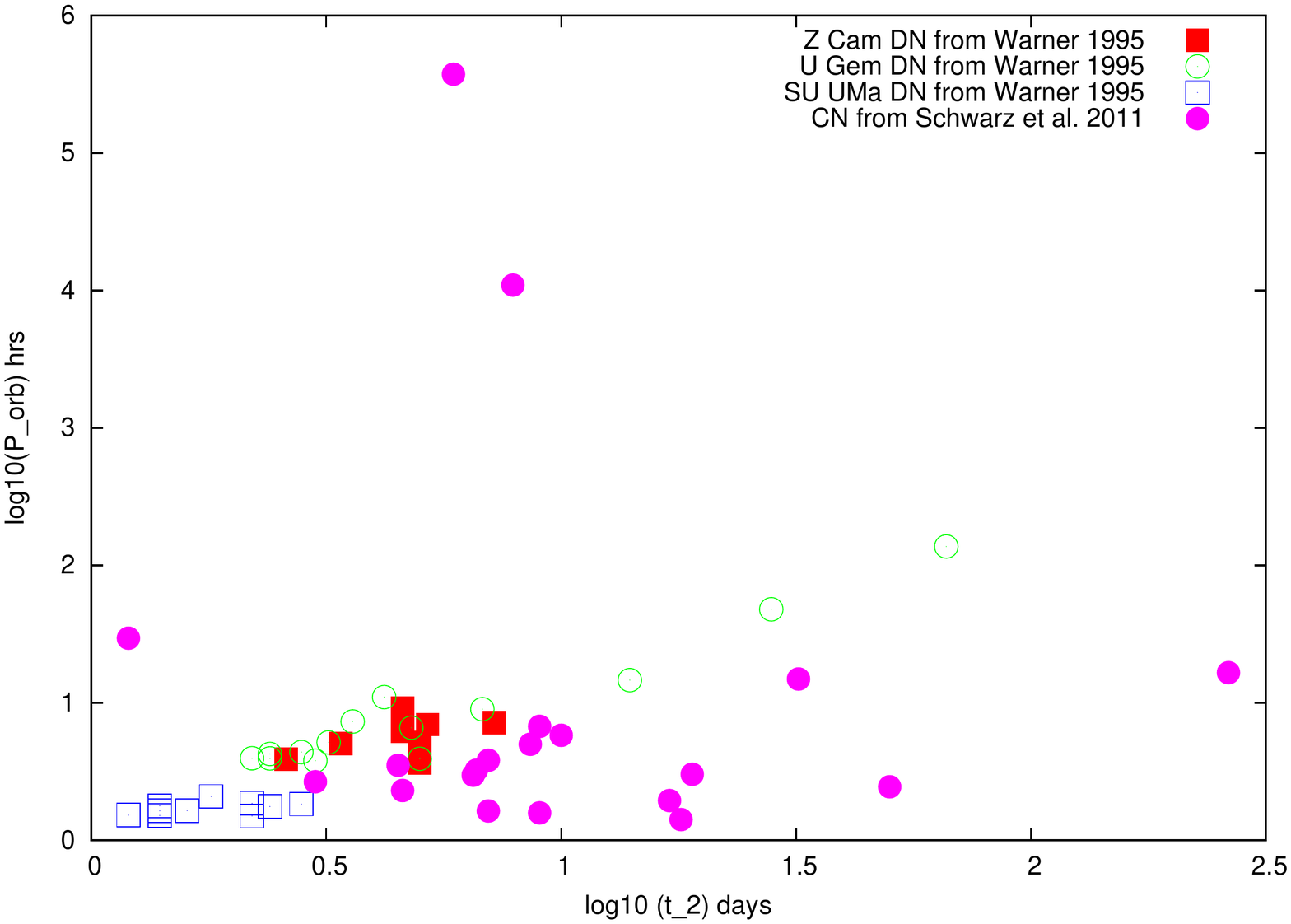}(b)
\caption{\small Figure show the orbital period $\rm P_{orb}\rightarrow t_2$.
(a) Data on dwarf novae are plotted.  A correlation is discernible especially for
the Z Cam and U Gem novae. Data 
taken from \citet{1995warner.book.....W}.  (b) Data on classical novae
are now included.  No correlation is noticeable. Data taken from \citet{2011ApJS..197...31S}. }
\label{fig4}
\end{figure}

\subsection{Dwarf novae}
\begin{figure}
\centering
\includegraphics[width=8.0cm]{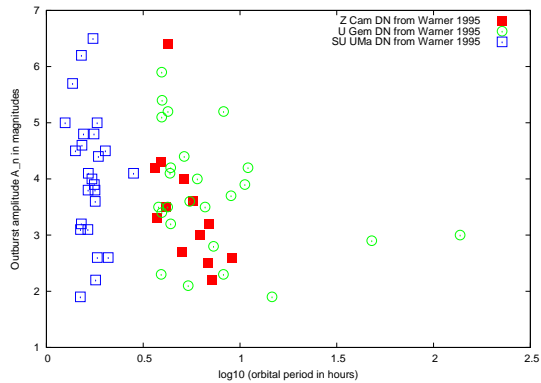}
\caption{\small The orbital period is plotted against the outburst amplitude for dwarf novae.
Data taken from \citet{1995warner.book.....W}.  A correlation is noted for the Z Cam DN.}
\label{fig5}
\end{figure}
Here we examine the applicability of the MMRD to dwarf novae.
In Figure \ref{fig3}(a), we have plotted the data on a few dwarf novae for which 
both the peak absolute magnitudes taken from \citet{1987MNRAS.227...23W} and $\rm t_2$
taken to be double of $\rm t_1$ listed in \citet{1995warner.book.....W} were available.   
The line shows the MMRD of Equation \ref{eqn15}.  The data on
dwarf novae do not follow the same trend as classical novae.  In fact the
data seems to suggest that there is no trend or that the trend is opposite to that of 
classical novae.  Hence an MMRD with a modified zero-point will not be applicable to 
dwarf novae.  In Figure \ref{fig3}(b),
the outburst amplitude $\rm A_n$ is plotted against $\rm t_2$ for a sample of
classical novae taken from \citet{2010AJ....140...34S}.  A trend similar to the MMRD 
in terms of the magnitude of the rate of change in brightness with $\rm t_2$ is noticeable
in $\rm A_n$  Thus, using the same magnitude of the coefficient of $\rm log_{10}t_2$ in
the MMRD of Equation \ref{eqn15} with a reversed sign, we 
find the best least squares fit to the data is  $\rm A_n=-2.16~log_{10}(t_2) + 13.68(\pm0.23)$
which is overplotted on the data.  The data on dwarf novae taken from 
Tables 3.1, 3.2 and 3.3 in \citet{1995warner.book.....W}
for the three types of dwarf novae namely the Z Cam, U Gem and SU UMa 
are also plotted in Figure \ref{fig3}(b).  In this case, the trend followed
by the dwarf novae is similar
to classical novae and hence just a modified zero-point is required. 
The best fit ($\rm A_n=-2.16~log_{10}(t_2) + 4.83(\pm0.16)$) is overplotted on the data points 
in the lower part of the figure.
The same rate of change in the outburst amplitude with $\rm t_2$ can be suggestive of a
common origin of the V band emission in both types of novae.  We recall
that the original calibration of the MMRD slope was from the fit to the
ejecta velocity and $\rm t_2$.  The difference between the behaviour in the two
panels for dwarf novae might be because both the peak and quiescence
absolute magnitudes of dwarf novae have been found to show a relation with the
orbital period of the binary \citep{1987MNRAS.227...23W}.  In Figure \ref{fig4}(a), the
orbital period is plotted against $\rm t_2$ and a correlation between the two quantities 
is noticeable so that the novae with longer orbital
periods have longer decline times.   This has already been pointed out 
by \citet{1995warner.book.....W}.  From the figure,
it is not clear if there is any correlation between $\rm P_{orb}$ and $\rm t_2$ for
the SU Uma class of dwarf novae.  In Figure \ref{fig4}(b), the data on classical
novae taken from \citet{2011ApJS..197...31S} have also been included.  No correlation between
orbital period and decline time is discernible for classical novae.  
In Figure \ref{fig5}, the orbital period of the binary and outburst amplitude for
dwarf novae are plotted.  Data have been taken from \citet{1995warner.book.....W}. 
Z Cam dwarf novae seem to show a correlation between $\rm P_{orb}$ and $\rm A_n$ whereas 
the U Gem and SU UMa do not (see Figure \ref{fig5}). 

The behaviour of the observables plotted in Figures \ref{fig3},\ref{fig4},\ref{fig5}
can give us important insight into the
physical processes which could be common or otherwise between classical and dwarf novae and can
help enhance our understanding of novae.  A quick analysis suggests that some of the differences
could be in the energetics of the outburst which are different for classical
and dwarf novae.  So for example, the outburst could be influenced by the companion
star in dwarf novae whereas the companion would play no role in an energetic
classical nova outburst.  However this needs to be examined, in detail, with available
data on these objects before any conclusions are drawn.

\subsection{Using the MMRD to estimate $\rm M_{V,0}$}
\begin{figure}[t]
\centering
\includegraphics[width=9.0cm]{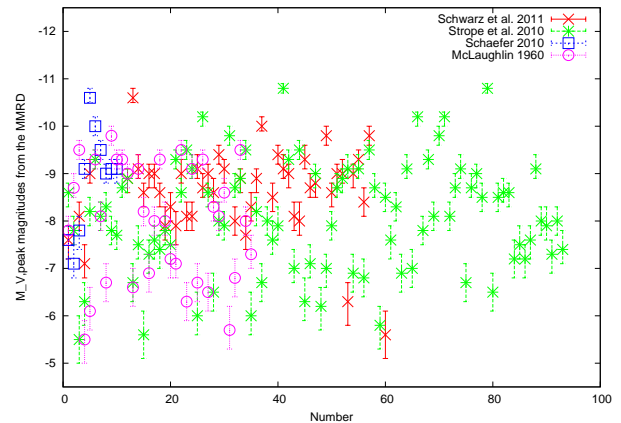}
\caption{\small Peak absolute magnitude determined from the MMRD using the observed $\rm t_2$ on 51 novae from
\citet{2011ApJS..197...31S}(crosses),  on 93 novae from \citet{2010AJ....140...34S}(stars), 10 recurrent
novae from \citet{2010ApJS..187..275S}(squares) and 35 novae from \citet{1960stat.conf..585M}
(circles).  The median value of the absolute
magnitude distribution is $-8.5$ magnitudes and the mean value is $-8.3$ magnitudes.  Most of the 
novae have a peak absolute magnitude between $-6$ and $-10$ magnitudes.}
\label{fig6}
\end{figure}

We used the MMRD in Equation \ref{eqn15} to estimate the peak absolute magnitudes of 
classical and recurrent novae for
which $\rm t_2$ is taken from literature and these are shown in
Figure \ref{fig6}.   Most of the novae have $\rm M_{V,0}$ between $-6$ and $-10$ magnitudes
which correspond to peak luminosities $10^{1.17} - 10^{5.17}$ L$_\odot$.
We have taken $\rm t_2$ for 51 novae from
\citet{2011ApJS..197...31S}, 93 novae from \citet{2010AJ....140...34S}, 10 recurrent
novae from \citet{2010ApJS..187..275S} and 35 novae from \citet{1960stat.conf..585M}.
The estimated $\rm M_{V,0}$ alongwith $1 \sigma$ uncertainties are listed for all 
these novae in Tables \ref{tab2},\ref{tab3},\ref{tab4},\ref{tab5} in the Appendix.  
Several novae are common to the different publications but sometimes listed with different
$\rm t_2$.  For example, 
there are 22 common novae in the catalogues by \citet{2010AJ....140...34S} and 
\citet{2011ApJS..197...31S} and owing to the different listed values of $\rm t_2$ for
some novae,  different $\rm M_{V,0}$ are determined by the MMRD for the same nova.  
We have estimated and listed $\rm M_{V,0}$ for the repeat entries also.  
It is important to homogenise all the data that are available
on novae so that the same values determined from light curves and spectra can be 
universally used and which can help reduce the uncertainties introduced into the
peak absolute magnitudes due to varying values of observables used by different astronomers.  
The median value of $\rm M_{V,0}$ of the $\sim160$ novae shown in Figure \ref{fig6} 
is $-8.5 $ magnitudes whereas the mean value is $ -8.3 $ magnitudes.  
We recall that the mean of the peak absolute magnitudes of 10 novae derived from 
different distance methods \citep{1960stat.conf..585M} are: 
from nebular expansion method: $-8.1$ magnitudes, from interstellar line intensity:  
$-7.6$ magnitudes, from Galactic 
rotation method: $-7.6$ magnitudes and from the trigonometric parallax:  $-7.3$ magnitudes. 
There is a large distribution ($-7.1$ to $-10.6$ magnitudes) in
the peak absolute  magnitudes estimated for recurrent novae (blue symbols in Figure \ref{fig6})
since the $\rm t_2$ of the ten recurrent novae range from 1.2 to 50 days 
\citep{2010ApJS..187..275S}.
Since recurrent novae are believed to be a possible progenitor channel for SN 1a, it
is intriguing that the recurrent novae show a relatively large range in their
$\rm t_2$ and hence $\rm M_{V,0}$.  The decline times and peak absolute magnitudes
of SN 1a are generally observed to be constrained to a narrow range, thus, making
them `standard candles'. 

\section{Distances to novae from MMRD}
Now that M$_{V,0}$ for a nova has been determined from the MMRD
as explained in the previous section, the distance to that nova can be calculated if the
peak apparent magnitude in the V band, m$_{V,0}$ of the nova is precisely known.
The distance modulus which is useful for quoting extragalactic distances is defined by:
\begin{equation}
\rm
m_{V,0} - M_{V,0} = 5 log_{10} (\frac{D}{10})~~ magnitudes
\label{eqn16}
\end{equation}
where D is the distance to the nova in parsecs, which is generally used for quoting
Galactic distances, will be: 
\begin{equation}
\rm
D = 10^{0.2(m_{V,0}-M_{V,0})+1} ~~ pc
\label{eqn17}
\end{equation}
If extinction $\rm A_V$ to the nova is known 
then including it will modify the apparent magnitude in Equation \ref{eqn17} to $\rm m_{V,0} - A_V$
which will reduce the distance estimate.  In the following sections, we estimate the distances
to the novae for which $\rm M_{V,0}$ has been determined from the MMRD of Equation \ref{eqn15}
and $\rm m_{V,0}$ is available.  We discuss the results on Galactic and extragalactic novae. 

\subsection{Galactic novae}
\label{sec-galnova}
Distances estimated from Equation \ref{eqn17} using the MMRD $\rm M_{V,0}$ and
$\rm m_{V,0}$ taken for
51 novae from \citet{2011ApJS..197...31S}, 93 novae from
\citet{2010AJ....140...34S}, 10 recurrent novae from \citet{2010ApJS..187..275S} and 35 novae from
\citet{1960stat.conf..585M} are shown in Figure \ref{fig7} and tabulated in 
Tables \ref{tab2},\ref{tab3},\ref{tab4},\ref{tab5} in the Appendix.
The three novae in LMC from Table \ref{tab2} are not included in the figure. 
Many Galactic novae appear to lie beyond 
20 kpc and two novae are even placed at a distance of 100 kpc! These large distances
for Galactic novae are clearly
unrealistic and indicate an error in $\rm M_{V,0}$ or $\rm m_{V,0}$ which is propagated
to the distance estimate where it is easily identifiable.  

\begin{figure}
\begin{center}
\includegraphics[width=8.7cm]{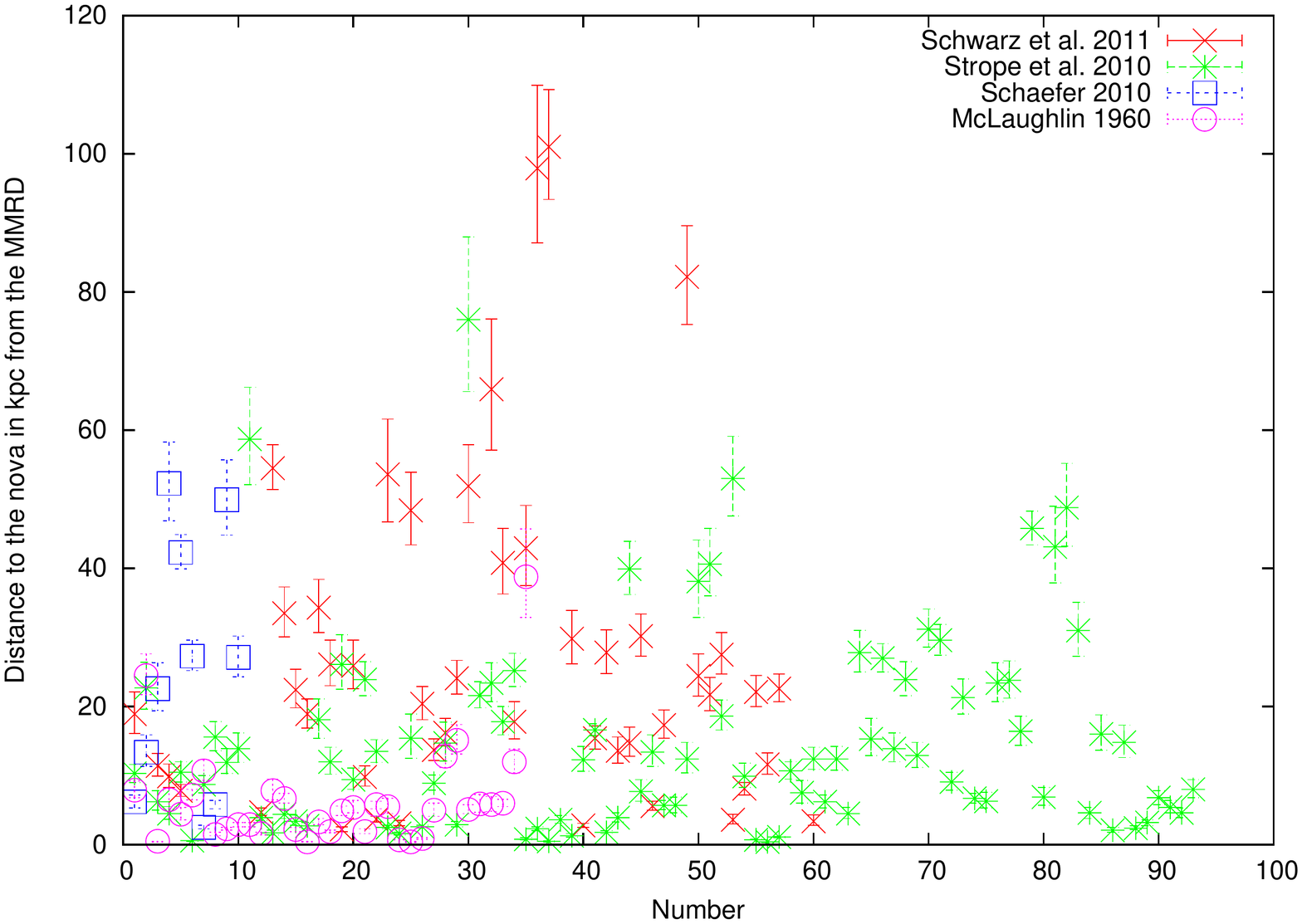}(a)
\includegraphics[width=8.7cm]{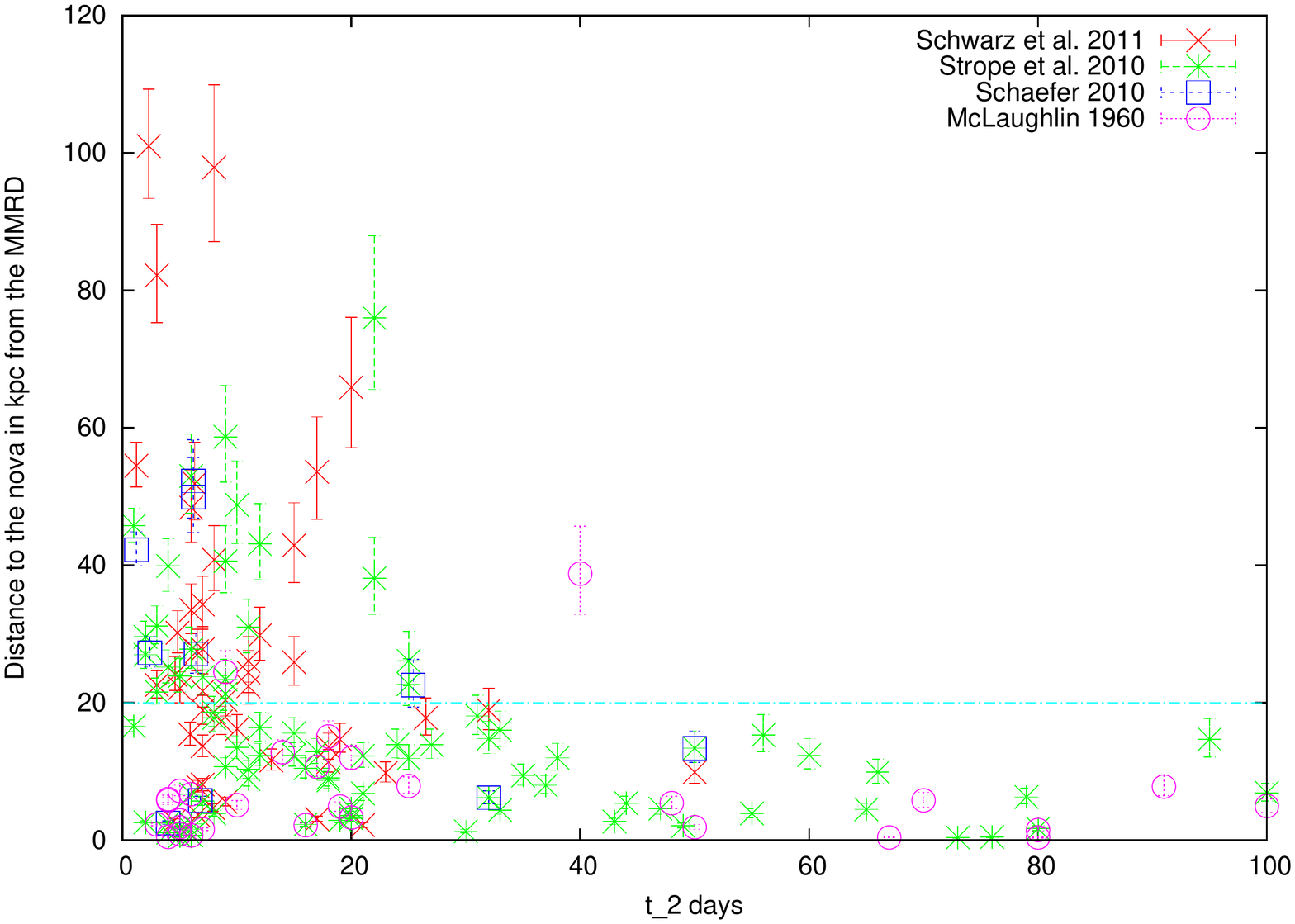}(b)
\caption{\small Distances estimated for a sample of Galactic novae for which $\rm M_{V,0}$ is
derived from the MMRD using their $\rm t_2$ listed in 
\citet{2011ApJS..197...31S,2010AJ....140...34S,2010ApJS..187..275S,1960stat.conf..585M}.
(a) All novae are plotted.  
(b) The distances are plotted as a function of $\rm t_2$ and only those
novae with $\rm t_2<100$ days are plotted for clarity.  Note the large distance estimates
for novae with $\rm t_2 < 25$ days.  }
\label{fig7}
\end{center}
\end{figure}
The uncertainty on the distance estimates plotted in 
Figure \ref{fig7} reflect the uncertainty in the  MMRD-determined peak absolute magnitudes 
and these appear to be fairly small indicating well-constrained distance estimates
especially for novae which have distances smaller than Galactic
dimensions.  The uncertainty on the peak absolute magnitude and hence distance, 
assuming $\rm m_{V,0}$ is precise, is smallest for
the fastest novae (see Figure \ref{fig2}(b)).  
For example, KT Eridani which has a $\rm t_2=6.6$ days, MMRD finds 
$\rm M_{V,0}=-9(-9.3,-8.8)$ magnitudes and if $\rm m_{V,0}=5.42$ magnitudes,
then it is located at a distance of
$7.8(8.7,7)$ kpc while V382 Vel which has a $\rm t_2=4.5$ days, MMRD finds
$\rm M_{V,0}=-9.4(-9.6,-9.2)$ and for $\rm m_{V,0}=2.85$ magnitudes, it is located
at a distance of $2.8(3.1,2.5)$ kpc (see Table \ref{tab2}). 
If the value of $\rm t_2$ is
correct to within a day or so then the value of $\rm M_{V,0}$ is likely to be correct 
for the novae other than the very fast ones since $\rm M_{V,0}$ is estimated from
the logarithm of $\rm t_2$.
Thus it is more likely that the distance errors result from an erroneous $\rm m_{V,0}$. 
$\rm m_{V,0}$ could be in error 
for fast novae if the peak of the optical light curve is missed i.e. the 
nova is detected post-maximum.  In this case, the large distances would have been estimated
predominantly for fast novae.  To check this, we plotted the distances estimated to these novae
against their $\rm t_2$ as shown in Figure \ref{fig7}(b).  
We note that all the three novae with distance estimated to be $>80$ kpc are 
fast with $\rm t_2 < 10$ days. 
Also most of the novae for which distances $>20$ kpc are estimated have a 
$\rm t_2<25$ days.  This, then, suggests that 
the recorded value of $\rm m_{V,0}$ could be offset from
the actual peak apparent magnitude due to the detection of the nova when it has already
started its decline.  We examine this further in section \ref{sec3.4}.
Interestingly, the larger-than-Galaxy distances ($> 20$ kpc) are estimated
for only 2/35 novae listed in
\citet{1960stat.conf..585M} as compared to 25/48 for novae listed in 
\citet{2011ApJS..197...31S}, 24/93 for novae listed in \citet{2010AJ....140...34S} and 
6/10 recurrent novae with data taken from \citet{2010ApJS..187..275S}.  
Such results have prompted suggestions in literature on limitations of the MMRD,
on the requirement of separate MMRD calibration 
for fast and slow novae; that MMRD is not applicable to recurrent novae.  
Such hasty conclusions are wrong and are arrived at before ruling out other more obvious reasons. 
The MMRD of Equation \ref{eqn15} plotted in Figure \ref{fig2} and the estimated 
$\rm M_{V,0}$ shown in Figure \ref{fig6}
strongly argue for the applicability of the same MMRD to both classical and recurrent novae
irrespective of the speed class as long as an accurate $\rm t_2$ is available.  We examine,
in detail, the possible reasons for the wrong distance estimates in section \ref{sec3.4}.

\subsection{Novae in M31}
\begin{figure}
\centering
\includegraphics[width=8.0cm]{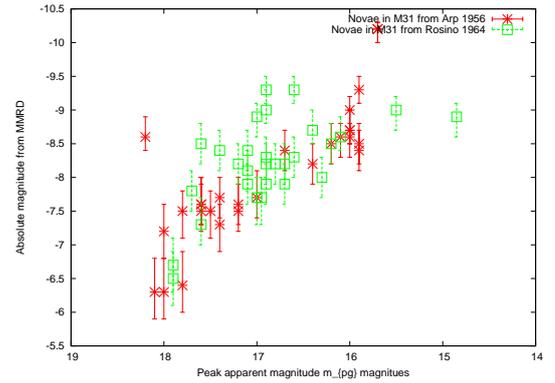}(a)
\includegraphics[width=8.0cm]{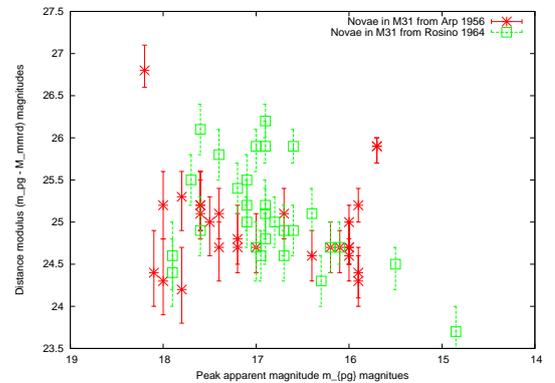}(b)
\includegraphics[width=8.0cm]{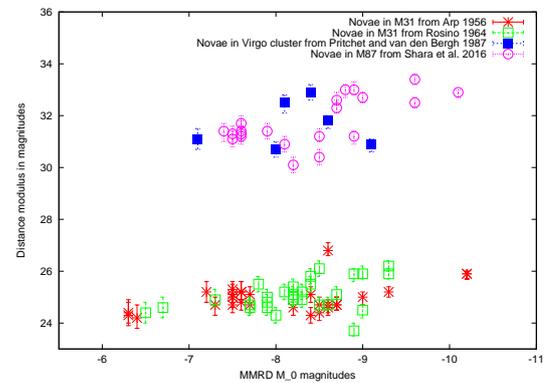}(c)
\caption{The $\rm M_{V,0}$ estimated using the MMRD relation
for novae in M31. $\rm t_2$ taken from \citet{1956AJ.....61...15A,1964AnAp...27..498R}.
(a) $\rm M_{V,0}$ is plotted versus the recorded peak apparent magnitude
taken from the same two references. (b) The distance modulus is plotted
versus the peak apparent magnitude. (c) The  
distance modulus is plotted versus the peak absolute magnitude for the novae in M31 and six novae
in  galaxies in the Virgo cluster listed in \citet{1987ApJ...318..507P} and 21 novae
in M87 from \citet{2016ApJS..227....1S}. }
\label{fig8}
\end{figure}
The MMRD in Equation \ref{eqn15} was used to estimate the peak absolute magnitudes of the novae
recorded in outburst in the neighbouring galaxy M31 and the distance modulus was estimated 
for the novae whose $\rm t_2$ and $\rm m_{V,0}$ are taken from 
\citet{1956AJ.....61...15A,1964AnAp...27..498R}.  The results
are plotted in Figure \ref{fig8} and tabulated in Table \ref{tab6} in the Appendix.
\citet{1956AJ.....61...15A} found that the novae in M31 and our Galaxy had a
similar distribution of peak absolute magnitudes and light curve characteristics.  We also recall that
\citet{1929ApJ....69..103H} had concluded from his study which detected
69 new novae in M31 that the mean light curve of novae in M31 was similar to that of
Galactic novae.  Moreover \citet{1929ApJ....69..103H} had estimated that the peak absolute magnitudes
of novae in M31 were distributed within $\sim 4$ magnitudes and that the rate of novae were 
about 30 per year in M31 which we note are similar to what we know for 
our Galaxy.  This would then imply that novae in
both galaxies share common physics and the same MMRD would apply to novae in M31 
and in our Galaxy.  However some subsequent studies in literature have hinted at 
an independent MMRD calibration for novae in M31.  We revisit this issue here using
our better calibrated MMRD. 
In Figure \ref{fig8}(a), $\rm M_{V,0}$ of novae estimated using the MMRD are plotted versus $\rm m_{V,0}$ 
and since the novae lie at almost the same distance from us,
the expected correlation of intrinsically bright novae also appearing 
brighter in the apparent magnitude is discernible.  
This is encouraging for the MMRD since $\rm M_{V,0}$ has been
estimated from the recorded $\rm t_2$ whereas $\rm m_{V,0}$ has been independently recorded.
The plot shows that the peak absolute magnitudes of most novae
in M31 lie between $-6$ and $-9.5$ magnitudes which is similar to novae in our Galaxy as
Arp and Hubble had already noted (see Figure \ref{fig6}).   
The outlier points such as intrinsically bright but
apparently faint or vice versa might indicate an error in 
$\rm  t_2$ or $\rm m_{V,0}$ or are novae for which the scatter on 
MMRD is larger due to our assumptions in Section \ref{sec-origin}. 
In Figure \ref{fig8}(b), the distance modulus is plotted which places
most of the novae between 24 and 26.8 magnitudes. The mean of
the distance modulus estimated for the 57 novae in M31 is 25.01 magnitudes. 
A quick analysis of the results for M31 shows
that there are three novae from Arp's catalogue which appear as the outliers in Figure \ref{fig8}.
One of the outlier points is a bright nova ($\rm M_{V,0}\sim-8.6$ magnitudes)
but is recorded to be fairly faint at maximum thus placing it a large distance
of 26.8 magnitudes.  From examining the light curve of this nova in Arp's paper (No 4), we find
that it was detected before optical maximum.  It is hence difficult to understand the cause
of such a large distance estimate in terms of observational quantities since both
$\rm m_{V,0}$ and $\rm t_2$ appear correct. 
We do not comment on it further here but keep it aside for possible future study. 
The remaining two outlier novae are estimated to be the intrinsically brightest
of the sample at $\rm M_{V,0} = -10.2$ magnitudes but do not appear bright enough in $\rm m_{V,0}$
so that they are placed at a large distance of 25.9 magnitudes.  
On examining the light curves of these two novae in Arp's paper (No 1,2), we note that for neither novae
which are very fast at $\rm t_2=2$ days, the peak
apparent magnitude has been observationally recorded.  $\rm m_{V,0}$ has been interpolated in the 
first case and extrapolated
in the second case and hence the possibility of these being underestimated cannot be ruled out. 
If these three novae are not included then the 25 novae in the Arp sample give a distance
modulus of 24.8 magnitudes to M31.
It would be premature and hasty to disbelieve the MMRD or cast doubts
on the relation based on such rare outlier points when it is found to work well with most novae. 
In fact, such outlier points should be examined in detail.  In Figure \ref{fig8}(c), the
distance modulus is plotted against $\rm M_{V,0}$ for the novae in M31 and Virgo cluster.
In both cases, we note that the scatter on the distance modulus estimated for faster i.e. brighter
novae is larger.   

\subsection{Novae in Virgo cluster galaxies}
\begin{figure}[t]
\centering
\includegraphics[width=8.0cm]{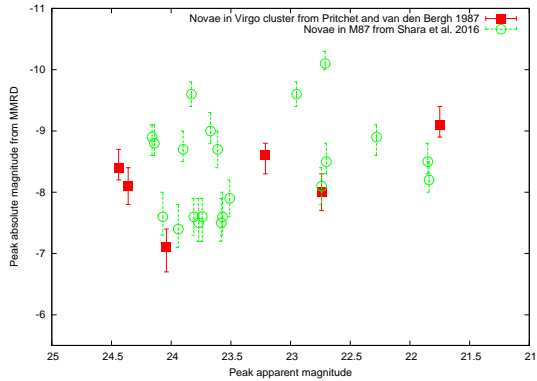}(a)
\includegraphics[width=8.0cm]{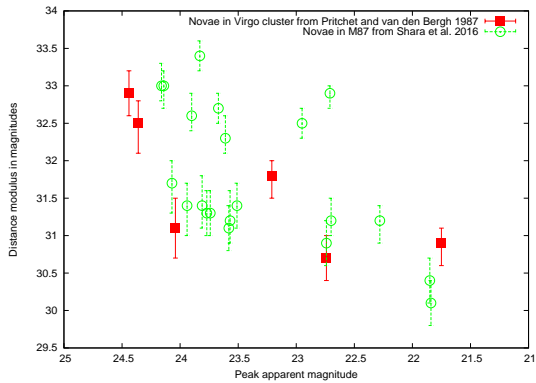}(b)
\caption{The $\rm M_{V,0}$, estimated using the MMRD relation
of novae in outburst in Virgo cluster with 
$\rm t_2$ taken from \citet{1987ApJ...318..507P,2016ApJS..227....1S}.
(a) $\rm M_{V,0}$ is plotted versus the recorded peak apparent magnitude. 
(b) The distance modulus is plotted versus the peak apparent magnitude.  There is a large
scatter in the distance modulus which appears to be due to observational constraints and not MMRD. }
\label{fig9}
\end{figure}

Using the recorded $\rm t_2$ in galaxies in Virgo cluster including M87,
and assuming that the same calibration of the MMRD applies to these novae, we
estimate their peak absolute magnitudes.   The MMRD $\rm M_{V,0}$  and distance modulus
estimated for these novae are plotted against their recorded peak apparent magnitude in
Figures \ref{fig8}(a),(b).  The results are listed in Table \ref{tab7} in the Appendix.
The data on $\rm t_2$ and $\rm m_{V,0}$ for these novae have been taken from 
\citet{1987ApJ...318..507P,2016ApJS..227....1S}.  
As in our Galaxy and M31, the peak absolute magnitudes range
from $-7$ to $-10$ magnitudes with the somewhat brighter lower limit possibly
being dictated by sensitivity owing to the distant nature of the cluster.  
However the trend shown by the novae in M31 (see Figure \ref{fig8}) 
of intrinsically bright novae also appearing brighter when their peak apparent magnitude is recorded,
is only faintly discernible in the novae in Virgo cluster (see Figure \ref{fig9}). 
The scatter in the data on novae in Virgo cluster is clearly larger.  
All the novae lie between distance modulus of 30 ($\sim 10$ Mpc) and 33.4 ($\sim 47.8$ Mpc)
magnitudes.
On examining the light curves of 21 novae in M87 presented in \citet{2016ApJS..227....1S}, we find
that the light curves of novae numbered 3,4,8,13,18,19,22  appear to be
insufficiently sampled and these as noted in the table in the paper,
are the faster novae in the sample.  The mean distance modulus to M87 after excluding
the aforementioned seven novae is $31.3$ magnitudes ($\sim 18.2$ Mpc).  
If all the 21 novae are used, 
then the distance modulus to M87 is estimated to be $31.76$ magnitudes ($\sim22.5$ Mpc). 
This distance modulus to M87 is very similar to that derived using other
distance estimators and we believe gives ample proof to the validity of the MMRD 
presented here for the novae in Virgo cluster.  \\

From the discussion presented in this section, 
it appears that a good way to approach the MMRD is to derive the best possible
calibration using high quality data on novae mainly in our Galaxy as has been done in this paper.
This well-calibrated MMRD can then be used to derive peak absolute magnitudes and
distances to nova outburst in our Galaxy and neighbouring galaxies. 
While it will work on most classical novae, it might give perplexing results for a few
novae like the one from Arp's list of novae in M31.  Such cases can then be
examined in detail. 
When outlier novae occur, an error in observed quantities should be suspected
before the MMRD is blamed.  In literature, there is, sometimes, a tendency to doubt the MMRD
before ruling out observational constraints and data limitations which can
confuse our understanding of novae instead of improving it. 
If one can find an easy explanation for the scatter and outliers in the MMRD
in observational uncertainty then that should be preferred over doubting the veracity
of the relation.

The MMRD thus emerges as a reliable estimator of the peak absolute luminosity
of the nova given its decline time and subsequently distance given its
peak apparent magnitude.  Moreover for extragalactic novae, it appears that the
distances are best estimated using relatively slow novae with sufficient
sampling which would seems to result in more accurate estimates of the peak apparent magnitudes
and $\rm t_2$.   The peak absolute magnitudes and distances determined from the
MMRD appear limited by observational uncertainties and we discuss this more in the next section. 

\subsection{Reasons for large MMRD distances}
\label{sec3.4}
As noted in section \ref{sec-galnova}, several Galactic novae are placed outside the Galaxy
when the distance to the nova is estimated using the MMRD-determined $\rm M_{V,0}$
and observed $\rm m_{V,0}$ (see Figure \ref{fig7}).  Either we hastily postulate that
these are a new population of novae which lie in the intragroup medium close to the Milky Way in
the local group of galaxies or we investigate possible observational errors that could
lead to such incorrect distance estimates.  
We believe that these novae are inside our Galaxy and that the
distance estimates are errorneous.  We begin by enumerating the possible causes of errors 
in the procedure which  which we then discuss in detail: 
(1) a calibration error in MMRD, 
(2) an incorrect $\rm t_2$,  
(3) an incorrect $\rm m_{V,0}$,
(4) a limitation of the MMRD due to the assumptions. 

\paragraph{1. Calibration error in MMRD:}
Such an error should lead to incorrect estimates of $\rm M_{V,0}$ for most novae
except the ones which coincidentally fall in the region of the MMRD common to
the erroneous MMRD and correct MMRD.   
Considering we have used data measured on 51 novae 
to determine the relation $\rm \Delta v \rightarrow t_2$ which finds resonance in the relation
derived by \citet{1960stat.conf..585M} builds confidence in our result regarding the
rate of change of the peak absolute magnitude with decline time.   Moreover, further
calibration has been done using the peak absolute magnitudes of novae determined
from independent methods.  The MMRD thus arrived at in Equation \ref{eqn15} and
shown in Figure \ref{fig2}(b) is found to fit the peak absolute magnitude
$\rightarrow t_2$ data taken from several carefully compiled
catalogues.   Moreover, the distance estimates to many novae using the MMRD $\rm M_{V,0}$ 
are well within Galactic dimensions.  Thus, we have several reasons to believe that the 
calibration of the MMRD presented here is fairly reliable and not responsible for the 
large distance estimates. 

\paragraph{2. Incorrect $\rm t_2$:} 
\label{sect-t2}
Since $\rm M_{V,0}$ is estimated from the logarithm of
$\rm t_2$, the effect of a small error in the decline time will not result in a large
error in $\rm M_{V,0}$ of a nova.   Only a large fractional error $\rm \Delta t_2 / t_2$
will translate to a large error in the peak absolute magnitude of a nova outburst.
We consider the effect of an error in $\rm t_2$ on fast novae such as U Sco  which has a recorded 
$\rm t_2=1.2$ days (Table \ref{tab4}) for which the MMRD predicts $\rm M_{V,0}=-10.6(-10.8,-10.5)$ 
magnitudes.  For example, if we incorrectly
recorded the $\rm t_2$ of U Sco to be 2.4 days then the MMRD will give $-10.0$ magnitudes
for the peak brightness and if the error was larger such that $\rm t_2=5$ days is
recorded, then the MMRD will predict $\rm M_{V,0}=-9.3$ magnitudes. 
If the peak apparent magnitude is correctly recorded for this nova, then this kind of error 
in which the peak absolute brightness is underestimated will
result in an incorrect smaller distance estimate to the nova. 
If on the other hand, the decline time is incorrectly recorded to be faster say
$\rm t_2 = 0.5$ days, then the MMRD gives $\rm M_{V,0}=-11.45$ magnitudes and for
a correctly recorded $\rm m_{V,0}$, such an error will result in the nova being placed at a
wrong larger distance to compensate for its recorded faintness.  

We now consider the effect of an error on $\rm t_2$ for slow novae.
Lets take the case of V723 Cas for which the decline time $\rm t_2=263$ days. 
The MMRD predicts that the peak brightness of the nova outburst would be 
$\rm M_{V,0}=-5.6(-6.1,-5.1)$
magnitudes.  Suppose the $\rm t_2$ is incorrectly recorded to be 270 days in which case the
MMRD will determine its peak brightness to be $\rm M_{V,0}=-5.55$ magnitudes which
is an error of only 0.05 magnitudes and is well within the MMRD uncertainty. 
Now if the $\rm t_2$ was incorrectly recorded to be 500 days,
the MMRD will give $\rm M_{V,0}=-4.97$ magnitudes.  
Thus, when an error in $\rm t_2$ leads to a larger value 
then it will lead to an underestimate of $\rm M_{V,0}$ and if
$\rm m_{V,0}$ has been correctly recorded, than this will result in a smaller-than-actual
distance to the nova being estimated.  On the other hand, if the error in $\rm t_2$
is in the other direction, ie if instead of 263 days, the $\rm t_2$ is recorded to
be $50$ days, then the MMRD will give a $\rm M_{V,0}=-7.1$ magnitudes.  
Such an error in $\rm t_2$ where $\rm M_{V,0}$ gets overestimated by the MMRD will
lead to an erroneously large distance estimate for the nova if $\rm m_{V,0}$ is correct. 
It appears that the effect of an error in $\rm t_2$ for slow novae is likely to be smaller 
than the errors in $\rm t_2$ of faster novae especially 
with the improved widespread observing facilities which would ensure that only a small
error in $\rm t_2$ occurs.  However large errors in $\rm t_2$ leading to 
incorrect distance estimates to slow novae
are not entirely impossible is demonstrated by the case of the
nova V2295 Oph which recorded an outburst in 1993 and has a recorded decline time 
of $\rm t_2=9$ days (Table \ref{tab5}).  The MMRD gives $\rm M_{V,0}=-8.7(-9.0,-8.5)$ magnitudes 
and a distance of 40.6 kpc, which has to be wrong, is estimated using the recorded $\rm m_{V,0}$. 
The light curve of this nova is flat-topped (see Figure \ref{fig13}(b)) and such
novae are generally slow with $\rm t_2>100$ days.  If we assume that V2295 Oph is of
the same nature and hence a slow nova then it has been
detected  well after it was at its peak brightness.  Hence
$\rm t_2$ has been grossly underestimated and its peak brightness has been overestimated.
This error in $\rm M_{V,0}$ has
then propagated into its distance estimate so that the faint $\rm m_{V,0}=9.3$ magnitudes which
was recorded is compensated by placing it at a distance of 40.6 kpc. 
If we assumed that the correct $\rm t_2$ was 100 days then $\rm M_{V,0} = -6.48$ magnitudes
and the distance to the nova would be 14 kpc which would be within Galactic dimensions. 
Thus, for V2295 Oph, a large error in $\rm t_2$ appears to be the reason for its large
distance estimate. 

\begin{figure}
\centering
\includegraphics[width=8cm]{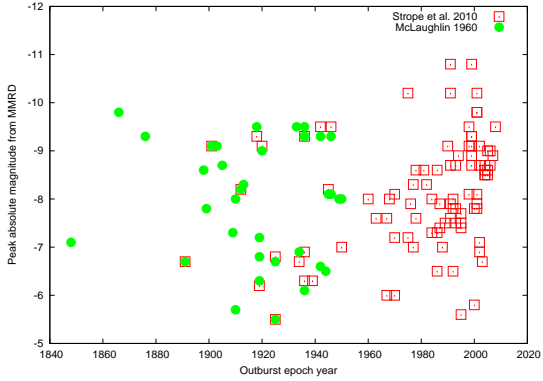}
\caption{\small 
The figure shows the distribution of $\rm M_{V,0}$ estimated by the MMRD against the
outburst epoch for the
novae listed in \citet{1960stat.conf..585M,2010AJ....140...34S}.  Note the
detection of several novae brighter than $-10$ magnitudes after 1960. }
\label{fig10}
\end{figure}

The above examples of fast and slow novae underline the logarithmic dependence
of $\rm M_{V,0}$ on $\rm t_2$ so that only large fractional errors on $\rm t_2$ have
a perceptible effect on $\rm M_{V,0}$ estimated from the MMRD and consequently 
distance to the nova.  Thus it appears that while an incorrect $\rm M_{V,0}$
from the MMRD might be responsible for the
unrealistic distances for a few novae, it is unlikely to explain
the same for so many novae.  In Figure \ref{fig10}, the $\rm M_{V,0}$ estimated
from the MMRD is plotted against the epoch of the nova outburst for the novae listed in
\citet{1960stat.conf..585M,2010AJ....140...34S}.  It is interesting to
note that several bright novae $\rm M_{V,0} < -10$ magnitudes have been detected 
since 1960 while none were detected before 1960 in the samples plotted here.

\paragraph{2. Incorrect $\rm m_{V,0}$:} 
\begin{figure}[t]
\centering
\includegraphics[width=8cm]{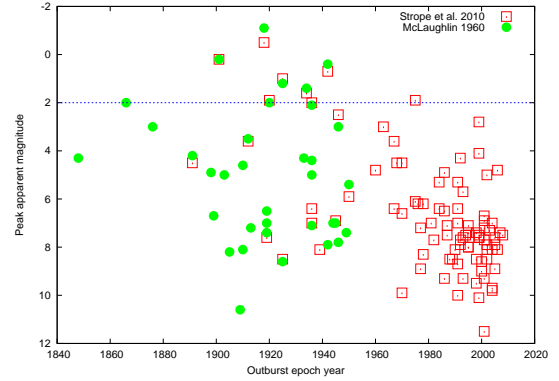}(a)
\includegraphics[width=8cm]{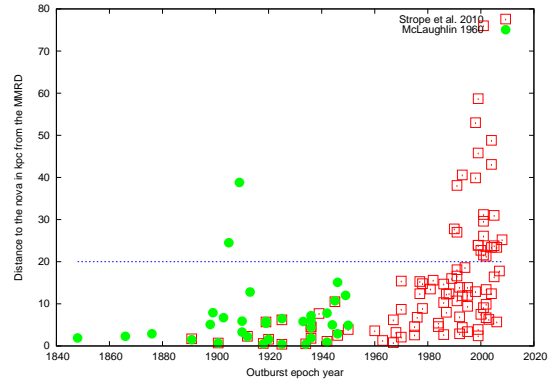}(b)
\caption{\small (a) The $\rm m_{V,0}$ is plotted with the outburst epoch of a nova from two catalogues 
to show that few novae with $\rm m_{V,0}<2$ have been detected in the last fifty years as compared
to 1900 to 1950. The horizontal line is drawn at 2 magnitudes.
(b) The MMRD distances are plotted versus the outburst epoch and shows a peculiar correlation.
Notice that post-1990 the distances to several novae are estimated to be $>20$ kpc and that the
scatter on distances is larger for novae which recorded an outburst post-1960.  
The horizontal line is plotted at 20 kpc.  }
\label{fig11}
\end{figure}
An error in the peak apparent magnitude can result from dust extinction or if the nova
is detected after optical maximum when the brightness is declining.  
Dust extinction will be most severe for the novae 
located in the Galactic plane especially close to the Galactic centre where it can be
as large as $2-3$ magnitudes.  This can introduce an error in the distance
estimate if left uncorrected.  An $\rm m_{V,0}$ left uncorrected for extinction 
will always lead to a larger distance estimate to novae for a given $\rm t_2$. 
This could be an important contributor to the wrong distances to several novae 
located towards the Galactic centre region. 

The second important and
as we show below a more frequent error in $\rm m_{V,0}$ arises from missing the optical
peak of the nova light curve.  Due to the rapid rise
to maximum, several novae are detected when they are already on the decline
after the optical peak.  In absence of any information on when the optical peak 
might have occurred, the best we can do is assume that the first detected point indicates the 
maximum apparent magnitude of the nova or if the detection of data points allow, one
can interpolate and use an appropriate value for the $\rm m_{V,0}$.  However in both cases, we
have to accept that there is a high probability of unknown errors in the values 
which will lead to the peak apparent brightness being underestimated.  
If a correct $\rm t_2$ is recorded, 
the detected faintness will be attributed to the nova being located at a larger distance. 
We also note that the detection of a fast nova after the optical peak might not result
in a large error in $\rm t_2$ but can cause a large error in the value of
the peak apparent brightness.  The magnitude of error in $\rm m_{V,0}$  will 
depend on how soon after the optical maximum the nova is detected and its speed. 
For example, if a nova is detected couple of days following optical peak then the
error on $\rm m_{V,0}$ will be larger for a fast nova then it will be for a slow nova.
Moreover as with
the extinction error, the incorrect estimate of $\rm m_{V,0}$ is always in the same
direction ie it always underestimates the peak apparent brightness of the nova so
that its peak appears fainter than it should based on its intrinsic brightness and actual distance.  
This means that when this incorrect $\rm m_{V,0}$ is combined with the MMRD $\rm M_{V,0}$
then the distance to the nova is always over-estimated.  The error on distance
is never in the other direction since it is not possible to record
a brighter apparent magnitude for a nova. 
Thus an error in $\rm m_{V,0}$ caused either by extinction or post-maximum detection
always leads to a faulty larger distance estimate. 

In Figure \ref{fig11}, we plot the $\rm m_{V,0}$ against the outburst epoch
for the novae listed in \citet{1960stat.conf..585M,2010AJ....140...34S}.  
Assuming these are representative samples, we note that
before 1960, several novae with apparent magnitudes brighter than 2 magnitudes have been
detected while only one such nova is present in the sample between 1960 and 2005. 
This paucity of novae with $\rm m_{V,0} < 2$ magnitudes is perplexing. 
Several fainter novae have been detected post-1960 as seen in Figure \ref{fig11}(a)
which can be explained by the improved sensitivity of modern optical telescopes.  
However if we compare this with Figure \ref{fig10}, we find that post-1960, a larger number of
brighter novae as estimated by the MMRD using the recorded $\rm t_2$ 
have also been discovered.  Combining these two results, it is
obvious that all the bright novae which have recorded fainter apparent magnitudes will be
placed at unreasonably large distances.  This is demonstrated in Figure \ref{fig11}(b) where the
distance to the same sample of novae is plotted against the outburst epoch.  
The data shows that almost half the novae detected
after 1990 (23/48) have MMRD distances estimated to be $> 20$ kpc and that the scatter
on the distance estimates of novae which have had an outburst post-1960 is larger.  We know of no
evidence to the existence of a different type of population of novae having been 
detected since 1990 and hence believe that such a result
suggests that for most of the novae outburst after 1990, the $\rm m_{V,0}$
are underestimated and have led to incorrect distance estimates.   
It could be that a large fraction of the
novae discovered post-1990 are located in the Galactic centre region and hence $\rm m_{V,0}$
is highly extincted or it could be that majority of these novae have been discovered 
post-maximum and hence the $\rm m_{V,0}$ have been underestimated.  The larger scatter
in the distance estimates of the novae discovered since 1960 as compared to the novae
outbursts which have been pre-1960 could be due to genuine detection of more distant novae
owing to better telescope sensitivities.  Whether this is the case or whether the uncertainty
is larger owing to some observational shortcomings can only be commented on when 
more information on these is gathered.  We find the peculiar behaviour exhibited
by the novae as a function of their outburst epoch intriguing and we believe
that understanding this should be useful in making further progress. 
 
To illustrate how easy it is to underestimate the optical peak brightness, 
especially in fast novae, the example of the recurrent nova T CrB ($\rm t_2=4$ days)
which recorded a peak apparent magnitude of 2.5 magnitudes
during its outburst in 1866 and a peak apparent magnitude
of 3.5 magnitudes in its 1946 outburst \citep{2010ApJS..187..275S}, is instructive.  
Since recurrent novae have shown similar
light curve behaviour in all outbursts, the peak apparent magnitude of 2.5 magnitudes obtained from
1866 is likely the best estimate of $\rm m_{V,0}$ for T CrB that 
we have.  This demonstrates how easily a one-magnitude 
error can be introduced in the peak apparent magnitude of a nova outburst especially
for classical novae for which this facility of recurrent outbursts allowing the 
correct peak apparent magnitude to be determined is not available. 
We can only ensure that all novae are detected before or near the pre-maximum halt
so their peaks are well-determined or have to contend with an unknown error in
the value of the peak apparent magnitude of a nova outburst and the corresponding error
in the distance estimate if estimated from its peak absolute magnitude.

To further demonstrate how an incorrect $\rm m_{V,0}$ is leading to the 
unrealistic distance estimates to novae, we reproduce some
light curves from \citet{2010AJ....140...34S} which these authors have mainly derived
from the large volumes of useful data collected by the American Association of 
Variable Star Observers (AAVSO) and augmented by other sources of data in literature,
if required. 

\begin{figure}[t]
\begin{center}
\includegraphics[width=7cm]{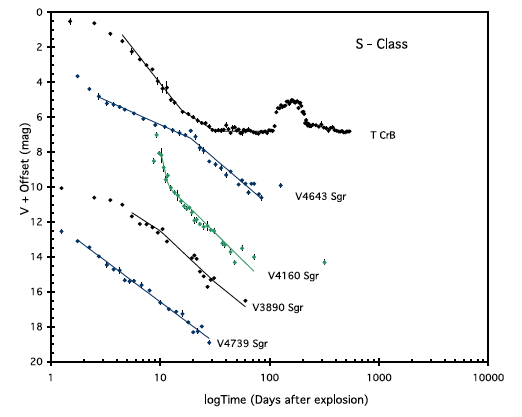}(a)
\includegraphics[width=7cm]{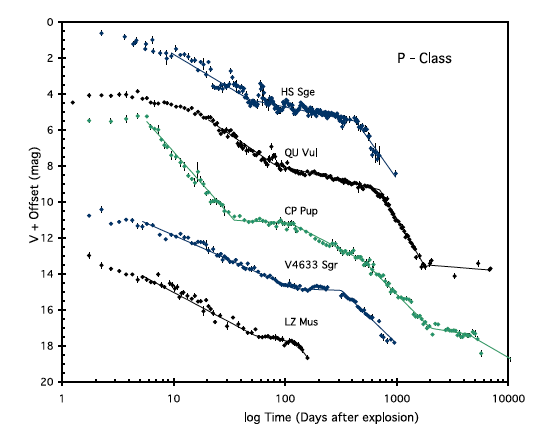}(b)
\caption{Light curves of 10 novae.  Figure copied from \citet{2010AJ....140...34S}. 
(a) Note the declining light curves from start 
for the novae (except T CrB) which can result in an incorrect $\rm m_{V,peak}$.
The MMRD distances to T CrB, V4643 Sgr, V4160 Sgr, V3890 Sgr, V4739 Sgr
are 2.5 kpc, 31.2 kpc, 27 kpc, 27.8 kpc, 29.6 kpc respectively. (b)
The light curves of QU Vul and CP Pup are flat when detected, rest show declining
light curves and possibly incorrect $\rm m_{V,0}$. 
The estimated distances to HS Sge, QU Vul, CP Pup, V4633 Sgr and LZ Mus are
12.4 kpc, 4.6 kpc, 1.1 kpc, 12.9 kpc and 39.9 kpc.  }
\label{fig12}
\end{center}
\end{figure}

We show light curves of 18 novae (10 in Figure \ref{fig12} and 8 in Figure \ref{fig13}) 
with the figures copied from \citet{2010AJ....140...34S}. 
In Figure \ref{fig12}(a), light curves of four novae are declining 
right from detection whereas T CrB shows some flattening at the top.  
In Figure \ref{fig12}(b), light curves of
QU Vel and CP Pup seem to be flattened at the top whereas the remaining three novae show a 
decline from detection.
We work with the obvious reasoning that the $\rm m_{V,0}$ is correctly estimated
from a light curve with a flat top whereas for the declining light curves, there is
a possibility of $\rm m_{V,0}$ being underestimated.  Thus, the distances to
novae with a flat top light curve should be smaller and well within Galactic dimensions
whereas for the declining light curves, the distance estimates could be larger. 
We find that the correlation between a flat topped light curve
and a small distance estimate well within Galactic dimensions is very good.  
Thus, of the novae in Figure \ref{fig12}(a),
T CrB is placed at a distance of 2.5 kpc whereas the
the other four novae with declining light curves are placed
beyond 20 kpc.  For the novae in Figure \ref{fig12}(b), the two flat-topped
novae QU Vul and CP Pup have the smallest distance estimates of 4.6 and 1.1 kpc of the 
five novae and the distances to HS Sge, V4633 Sgr and LZ Mus are estimated to
be 12.4, 12.9 and 39.9 kpc.  In Figure \ref{fig13}, all the eight novae show
flat-topped light curves and hence a correct $\rm m_{V,0}$ should be available for all of them.   
The distance estimated from the MMRD $\rm M_{V,0}$ and $\rm m_{V,0}$ 
to seven of these novae is within 10 kpc while one of them, 
V2295 Oph is placed at an unreasonable distance of 40.6 kpc.  Since it has a flat-topped
light curve, its $\rm m_{V,0}$ should be correct. 
We have already discussed this nova earlier in this section as an example of a large
error in $\rm t_2$.  We also add here that the $\rm t_2$ for the other
three flat-topped novae in  Figure \ref{fig13}(b) ie DO Aql, V849 Oph and BT Mon are
295 days, 140 days and 118 days \citep{2010AJ....140...34S} and the
the MMRD-determined $\rm M_{V,0}$ are $-5.5, -6.2, -6.3$ respectively. 
We are convinced that the large distance estimate to V2295 Oph is a result of a large
error in its estimated $\rm t_2$ of 9 days which has 
overestimated its $\rm M_{V,0}$ of $\rm -8.7$ magnitudes
so that even with a correct $\rm m_{V,0}$, the nova has been placed at a large distance.  
Obviously such cases, where a slow flat-topped nova is not detected for
several tens of days, will be rarer compared to errors in $\rm m_{V,0}$ due to detection 
of a nova on its decline.
Out of the eighteen novae light curves discussed here, the distances estimated
to six novae are greater than 20 kpc and five of these novae are 
detected on the decline.  The erroneous distance estimates for five of these novae 
can then be attributed to an incorrect $\rm m_{V,0}$ while one is traced to an incorrect $\rm t_2$.  

Several studies have likewise pointed out that the peak apparent magnitudes 
of novae are likely in error.  Here, we have tried to explain how serious the effect 
of the uncertainty can be when
the MMRD $\rm M_{V,0}$ and $\rm m_{V,0}$ are combined to estimate the distance to a nova.  
Moreover it is easier to identify a large  error in one of the observed quantity 
($\rm m_{V,0}$ or $\rm t_2$) when the distance estimate is combined with the light curve.
Importantly, the MMRD cannot be blamed for the erroneous distance estimates. 
Thus, we end this part after convincingly demonstrating that an error in
the peak apparent magnitude is the most frequent cause of erroneous distance
estimates to novae from the MMRD $\rm M_{V,0}$.  We should use the MMRD only when
a reliable estimate of $\rm m_{V,0}$ is available - in other words, the 
utility of the MMRD is maximal and distances reliable only when used on novae 
which are detected before their brightness begins to decline.

\begin{figure}[t]
\begin{center}
\includegraphics[width=7cm]{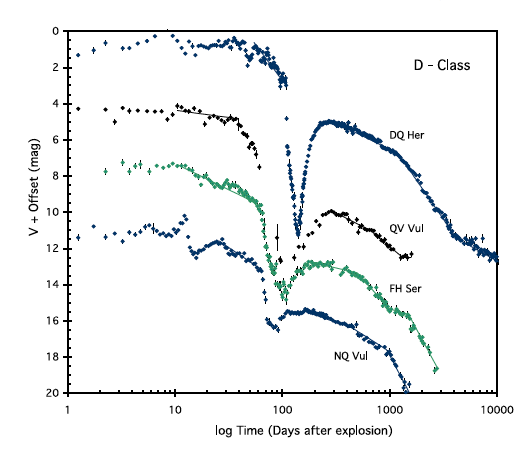}(a)
\includegraphics[width=7cm]{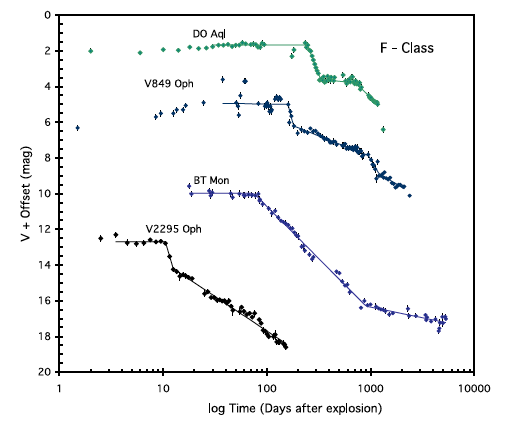}(b)
\caption{Light curves of eight novae. Figure copied from \citet{2010AJ....140...34S}.
(a) The light curves of all the novae are flat-topped indicating that 
$\rm m_{V,0}$ is correct.  Using the $\rm M_{V,0}$ from the MMRD, 
the distances to DQ Her, QV Vul, FH Ser and NQ Vul are 0.5 kpc, 8 kpc, 2.1 kpc and
6.8 kpc.  (b) The light curves of all the novae are flat-topped 
so $\rm m_{V,0}$ will be correct.
The estimated MMRD distances to DO Aql, V849 Oph, BT Mon and V2295 Oph are 6.2 kpc, 5.7 kpc,
7.7 kpc and 40.6 kpc.  We discuss the case of V2295 Oph in the paper
and find that its decline time $\rm t_2$ is in error.}
\label{fig13}
\end{center}
\end{figure}

\paragraph{Limitation of the MMRD:} 
As mentioned earlier and seen from Figure \ref{fig2}, the same MMRD relation 
(Equation \ref{eqn15}) is capable of predicting the peak absolute magnitudes of all kinds of 
Galactic novae: fast, slow, recurrent, classical and extragalactic novae.  
As demonstrated in the previous points, the distance estimates are mainly limited by errors on 
either the peak amplitude $\rm m_{V,0}$  which is often difficult to catch due to the extremely 
fast nature of the
rise in nova brightness and occasionally $\rm t_2$.  Thus, as stated at the beginning of
this section, till we are able to remove these shortcomings on the observed quantities,
it is premature to comment on the limitations of the MMRD.  The MMRD 
peak absolute magnitudes of most novae lie between $-6$ and $-10$ magnitudes (see
Figure \ref{fig6}) ie novae radiate at sub-Eddington to Eddington luminosities at its peak.   
The life-luminosity and life-expansion relations are derived 
assuming there is an instantaneous injection of the excess energy to the system.
These relations might give incorrect $\rm M_{V,0}$ if used for novae which show
rapid multiple peaks like Nova Herculis \citep{1936PASP...48..191Z}. 
This might especially be the case if the successive peaks occur within the timeframe of $\rm t_2$.  
Thus, if there is continuous injection of energy or several energy injection episodes
in a nova then it is possible that the MMRD will not be able to give a correct
estimate of its peak luminosity and this would constitute a limitation of 
the MMRD.  However whether this does happen in novae needs to be verified on 
actual data once the observational data are made error-free.  
Another limitation might arise from the assumptions that we made in
section \ref{sec-origin} while deriving the MMRD mainly that (1) the mass of the 
ejecta $\rm m_{ej}$ and (2) the maximum radiated energy $\rm E_{rad}$ are constant for all novae.   
Although the nova-to-nova differences in these
quantities might be small compared to other physical quantities such as $\rm v_{ej}$, 
these quantities cannot be constant for all novae.  Thus the range in these quantities for
novae of the same $\rm t_2$ can result in a scatter on the MMRD predicted
$\rm M_{V,0}$.  However when the observational uncertainties have been minimized, it might 
even be possible to probe the range of $\rm m_{ej}$ and $\rm E_{rad}$ 
involved in nova outbursts.   However as mentioned earlier, till we 
have rigourously shown that the distances and peak absolute magnitudes are not
limited by uncertainties on observational data, it can serve no purpose
to analyse or comment on these limitations of MMRD.
  
Thus at the end of this section, we conclude that the MMRD is a reliable
estimator of peak absolute magnitudes of novae provided its calibration is carefully
done and $\rm t_2$ is reliable.  It is a reliable distance estimator provided
$\rm m_{V,0}$ are accurate.  Currently, the  MMRD distances to novae are
limited by the accuracy of the nova observables, predominantly $\rm m_{V,0}$ 
and not by the uncertainties on the MMRD.  

\section{Intersection of novae light curves}
Since there are novae of different speed classes, there is a possibility that their 
absolute magnitude light curves might intersect on a particular day after maximum.
Knowing the expected value of the absolute magnitude of novae on that particular
day would be useful in estimating the distance to the nova.  Recognising the potential
of such an occurrence, \citet{1955Obs....75..170B} examined the light curves
of 11 novae of several speed classes and found that their absolute magnitudes showed 
minimum scatter about 15 days post-maximum.  They estimated a mean absolute
magnitude of $-5.2\pm0.1$ magnitudes on day 15.  Here we revisit the estimate using
data available on a larger number of novae.  We used data on eight novae from
\citet{1985ApJ...292...90C} and 28 novae from \citet{2000AJ....120.2007D} for 
which $\rm t_2$, $\rm M_{V,0}$ and 
$\rm M_{V,15}$ are made available in the papers.  The $\rm t_2$ and $\rm M_{V,15}$
listed in these papers are listed in columns 3,4 of Table \ref{tab1}. 
There are six common novae in the two lists.  As can be seen from Table \ref{tab1},
the $\rm M_{V,15}$ listed in \citet{1985ApJ...292...90C} range from 
$-4.8$ to $-8.85$ magnitudes and range from $-4.8$ to $-7.4$ magnitudes as
listed in \citet{2000AJ....120.2007D}.  This indicates that there
exists a large dispersion in $\rm M_{V,15}$ and we explore the possibility of
reducing it so that its usage as a distance diagnostic can be increased.  
Moreover we can use our
improved MMRD calibration to get a more reliable estimate of $\rm M_{V,15}$.
We proceed as follows.

\begin{table}
\centering
\small
\caption{\small The absolute magnitudes of novae on day 15. The first eight entries are from
\citet{1985ApJ...292...90C} and the next 28 are from \citet{2000AJ....120.2007D}. $\rm dm_{15}$
indicates the change in the nova brightness in magnitudes 15 days after peak brightness.
In the last two columns, the $\rm M_{V,0}$ and $\rm M_{V,15}$ ($\rm M_{V,0}+dm_{15}$)
estimated using the $\rm t_2$ listed in column 3 and our MMRD calibration are tabulated.  }
\begin{tabular}{l|l|c|l|l|l|l}
\hline
& {\bf Nova}  & \multicolumn{3}{c|}{\bf C 1985} & \multicolumn{2}{c}{\bf This MMRD} \\
& & { $\bf t_2$} & { $\bf M_{15}$} & {$\bf dm_{15}$} & {$\bf M_{0}$} &  { $\bf M_{15}$}  \\
& &     days     &    mag  & mag      & mag      & mag \\
\hline
1   &    V1229 Aql   &    18.0 & -4.8  & 1.8 &   -8.1  &    -6.3    \\
2   &    V500 Aql    &    20.0 & -8.85 & 1.5 & -8.0  &    -6.5  \\
3   &    V1500 Cyg   &    2.4  & -5.1  & 4.15 & -10.0 &    -5.8   \\
4   &    V446 Her    &    5.0  & -5.55 & 3.15 & -9.3  &    -6.1   \\
5   &    V533 Her    &    26.0 & -6.6  & 1.1 & -7.7  &    -6.6   \\
6   &    DK Lac      &    19.0 & -7.35 & 2.0 & -8.0  &    -6.0  \\
7   &    XX Tau      &    24.0 & -6.75 & 1.3 & -7.8  &    -6.5  \\
8   &    LV Vul      &    21.0 & -5.25 & 1.5 & -7.9  &    -6.4   \\
\hline
&  {\bf Nova} & \multicolumn{3}{c|}{\bf D \& D 2000 } & \multicolumn{2}{c}{\bf This MMRD} \\
& & {$\bf t_2$} & {$\bf M_{15}$} & {$\bf dm_{15}$} & $\bf M_{0}$  & {$\bf M_{15}$} \\
& &     days     &    mag   & mag     & mag      & mag \\
\hline
1   &    V500 Aql       &  17.0 & -6.8 & 2.2 & -8.1    &   -5.9       \\
2   &    V603 Aql       &  4.0  & -4.9 & 4.0 & -9.5    &   -5.5         \\
3   &    V1229 Aql      &  20.0 & -4.9 & 1.8 & -8.0    &   -6.2         \\
4   &    T Aur          &  45.0 & -6.2 & 0.8 & -7.2    &   -6.4        \\
5   &    V842 Cen       &  35.0 & -6.5 & 0.9 & -7.5    &   -6.6       \\
6   &    V450 Cyg       &  88.0 & -5.9 & 0.9 & -6.6    &   -5.7      \\
7   &    V476 Cyg       &  6.0  & -6.9 & 3.0 & -9.1    &   -6.1     \\
8   &    V1500 Cyg      &  2.4  & -5.5 & 5.2 & -10.0   &   -5.8    \\
9   &    V1819 Cyg      &  37.0 & -5.6 & 1.2 & -7.4    &   -6.2      \\
10  &    V1974 Cyg      &  17.0 & -6.4 & 1.6 & -8.1    &   -6.5        \\
11  &    HR Del         &  172.0& -4.8 & 1.3 & -6.0    &   -4.7         \\
12  &    DQ Her         &  39.0 & -6.55& 0.95& -7.4   &    -6.45      \\
13  &    V446 Her       &  7.0  & -6.5 & 3.4 & -9.0    &   -5.6       \\
14  &    V533 Her       &  22.0 & -6.1 & 1.4 & -7.9    &   -6.5      \\
15  &    CP Lac         &  5.3  & -5.7 & 3.6 & -9.2    &   -5.6     \\
16  &    DK Lac         &  11.0 & -7.4 & 2.4 & -8.6    &   -6.2    \\
17  &    GK Per         &  7.0  & -5.9 & 3.1 & -9.0    &   -5.9   \\
18  &    RR Pic         &  20.0 & -6.0 & 1.8 & -8.0    &   -6.2  \\
19  &    CP Pup         &  6.0  & -6.1 & 4.6 & -9.1    &   -4.5    \\
20  &    V351 Pup       &  10.0 & -6.1 & 2.3 & -8.6    &   -6.3      \\
21  &    FH Ser         &  42.0 & -6.45& 0.85 & -7.3   &    -6.45      \\
22  &    XX Tau         &  24.0 & -7.0 & 1.6 & -7.8    &   -6.2        \\
23  &    RW UMi         &  48.0 & -7.3 & 0.8 & -7.2    &   -6.4       \\
24  &    LV Vul         &  21.0 & -5.85& 1.15 & -7.9   &    -6.75    \\
25  &    NQ Vul         &  23.0 & -4.85& 1.25&  -7.9   &    -6.65   \\
26  &    PW Vul         &  82.0 & -5.3 & 1.4 & -6.7    &   -5.3    \\
27  &    QU Vul         &  22.0 & -6.35& 1.15 & -7.9   &    -6.75    \\
28  &    QV Vul         &  50.0 & -5.45& 0.95 & -7.1   &    -6.15       \\
\hline
\end{tabular}
\label{tab1}
\end{table}

\begin{figure}[t]
\centering
\includegraphics[width=8.0cm]{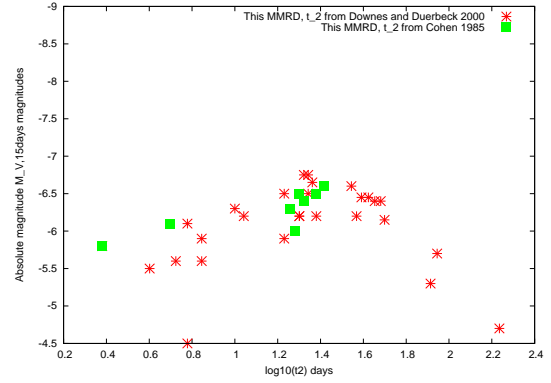}(a)
\includegraphics[width=8.0cm]{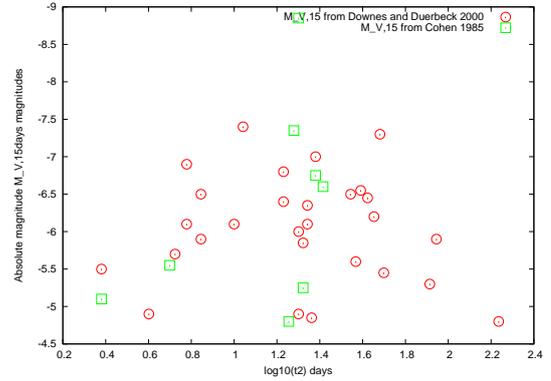}(b)
\caption{\small Absolute magnitude on day 15 after outburst maximum versus $\rm t_2$.  
(a) $\rm M_{V,0}$ estimated from the MMRD in Equation \ref{eqn15} and $\rm M_{V,15}$ estimated
from this using dm$_{15}$ from column 5 in Table \ref{tab1} is plotted versus $\rm t_2$.
A trend is discernible in the plot followed by all novae except
the nova plotted at $-4.5$ magnitudes and $\rm log_{10}t_2\sim0.8$.  
(b) $M_{V,15}$ for the same novae taken from \citet{1985ApJ...292...90C}(green boxes) and 
\citet{2000AJ....120.2007D} (red circles) are plotted.  Notice the comparatively larger 
scatter in this plot.  } 
\label{fig14}
\end{figure}

\begin{figure}[t]
\centering
\includegraphics[width=8.0cm]{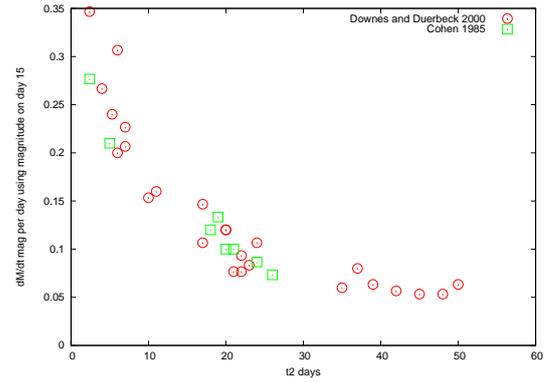}(a)
\includegraphics[width=8.0cm]{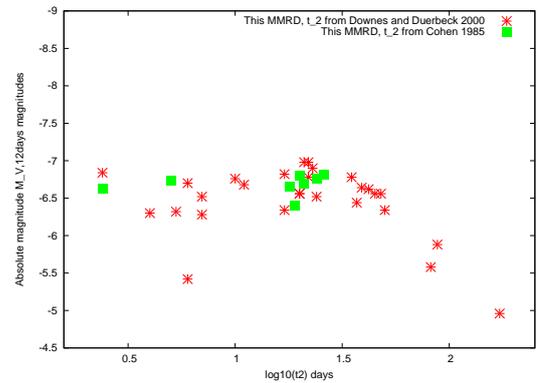}(b)
\caption{\small (a) The average rate of change in the brightness of novae per day estimated from
$\rm dm_{15}$ ie $\rm dm_{15}/15$ is plotted against $\rm t_2$. 
(b) The expected absolute magnitudes on day 12 estimated from $\rm M_{V,15}$ and
$\rm dm_{15}/15$.  The scatter on the mean value of
the absolute magnitude expected on day 12 seems less than on day 15 and the systematic variation
has been removed. The mean value of $\rm M_{V,12}$ is $-6.616\pm0.043$ magnitudes.  }
\label{fig15}
\end{figure}

We used the $\rm M_{V,0}$ and $\rm M_{V,15}$ listed in the papers to estimate the
difference in the amplitude of the star in 15 days and 
this is listed under $\rm dm_{15}$ in column 5 of Table \ref{tab1}.    
We also estimated $\rm M_{V,0}$ from the MMRD relation in
Equation \ref{eqn15} using the $\rm t_2$ from the papers and these are listed in column
6 of Table \ref{tab1}.  We, then, estimated $\rm M_{V,15}$ using our MMRD $\rm M_{V,0}$
and $\rm dm_{15}$ and these are listed 
in column 7 of Table \ref{tab1}.  The $\rm M_{V,15}$ estimated in this way lies
between $-4.5$ and $-6.8$ magnitudes for the 36 datapoints.  From comparing columns
4 and 7 of Table \ref{tab1}, we note that the two estimates of $\rm M_{V,15}$ - one
from the papers and other estimated by us using the MMRD $\rm M_{V,0}$ 
differ for several novae.  
In Figure \ref{fig14}(a), we plot the $\rm M_{V,15}$ estimated from our MMRD-determined
$\rm M_{V,0}$ and in (b) the $\rm M_{V,15}$ listed in 
\citet{1985ApJ...292...90C,2000AJ....120.2007D} with decline time.
An obvious difference between the two plots is the lower scatter on
$\rm M_{V,15}$ for a given $\rm t_2$ in Figure \ref{fig14}(a) as compared to (b). 
Moreover, a trend is discernible in (a) such that  $\rm M_{V,15}$ is fainter for fast
novae, increases for not-so-fast novae and again drops for slow novae. 
We believe that the more accurate estimates of $\rm M_{V,0}$ from our 
well-calibrated MMRD has allowed this behaviour of novae in the 
$\rm M_{V,15}-log_{10}t_2$ plane to be recognised.  An analysis of Figure \ref{fig14}(a)
indicates that since fast novae fade rapidly, their brightness has dropped to $\sim -5.5$ 
magnitudes by day 15 while the somewhat slower novae have faded to $\sim -6.5$ magnitudes 
and the really slow novae show the smallest change in their peak brightness in 15 days
and are already fainter to begin with.  Moreover, another inference which emerges from 
this figure is that the scatter could be reduced and a better estimate of the near-constant
absolute magnitude can be obtained, if we could remove this trend.  In other words, 
the absolute magnitude of novae of different speed classes might show lower scatter on a day 
other than day 15.  
To explore removing the systematic variation in Figure \ref{fig14}(a), we estimated
the mean rate of change in the absolute magnitude over 15 days after optical peak
i.e. $\rm dm_{15}/15$.  This is plotted in Figure \ref{fig15}(a) for the 36 datapoints.   
We then used this rate to estimate the absolute magnitudes of the novae on days 14, 13 and 12 
following outburst to check if the absolute magnitudes matched better on any of these days.   
Figure \ref{fig15}(b) shows the expected
absolute magnitudes on day 12 after outburst estimated as above.  The systematic trend has
been removed and $\rm M_{V,12}$ of 32/36 datapoints lie between $-6.2$ and $-7.0$ magnitudes. 
For comparison, the $\rm M_{V,15}$ that we had estimated was 
between $-5.4$ and $-6.8$ magnitudes for the same 32/36 points.
Clearly, the scatter on the absolute magnitude on day 12 is less than on day 15. 
The mean value of the expected absolute magnitude on day 12 using data on 24 novae is: 
\begin{center}
$\rm \bf M_{V,12} = -6.616\pm0.043~~magnitudes$ \\ 
\end{center}
We have excluded data on four novae while estimating the mean - the three slowest novae and 
the outlier near $\rm log_{10}t_2\sim0.8$ from the 28 novae listed in \citet{2000AJ....120.2007D}.
This is because the three slowest novae are fainter than or close to $-6.616$ magnitudes at
maximum and hence do not adhere to this relation which is also obvious
from Figure \ref{fig15}(b).  The fourth outlier nova is clearly an exception which 
needs to be examined further and hence has been excluded from the mean. 
It is important to mention that this $\rm M_{V,12}$ is expected to be
valid only for novae whose $\rm t_2$ lie betwen 2.4 days and 86 days.  
This method will obviously not work for novae fainter than $-6.616$ magnitudes 
i.e. very slow novae with $\rm t_2 > 86$ days.  
For comparison, the mean $\rm M_{V,15}$ estimated from the data on the same 24 novae
listed in column 7 of Table \ref{tab1} and shown in Figure \ref{fig14}(a) 
is $\rm -6.22 \pm 0.07$ magnitudes and is also
applicable to novae with $\rm t_2$ between 2.4 and 86 days.  

Thus, $\rm M_{V,12}$ provides a robust
alternative to $\rm M_{V,0}$ for estimating the distance to novae 
with $\rm 2.4 \le t_2 \le 86$ days, since it
requires $\rm m_{V,12}$ which might be easier to record with higher accuracy 
than is $\rm m_{V,0}$ for these novae.  
For the slowest novae ($\rm t_2 > 86$ days) which fade slowly, $\rm m_{V,0}$ can be 
accurately recorded and
the distance can be easily estimated from the $\rm M_{V,0}$ determined from the MMRD.

Remarkably, a better calibrated MMRD has enabled the determination of a better day after
outburst maximum when light curves of novae belonging to several speed classes 
intersect and a better
estimate of $\rm M_{V,12}$.  This, then, also provides strong evidence to the
validity of and the better calibration of the MMRD presented in the paper.  
$\rm M_{V,0}$ and $\rm M_{V,12}$ appear to be the best distance estimators to most novae which 
allow us to calculate the distance soon after the nova outburst. 
We end by listing the $\rm M_{V,15}$ from literature:
\begin{enumerate}
\item Mean $\rm M_{V,15} = -5.2\pm0.1$ magnitudes which was found to
be applicable for novae in our Galaxy, LMC and M31 by \citet{1955Obs....75..170B}.
\item Mean $\rm M_{V,15} = -5.6\pm0.45$ magnitudes by \citet{1985ApJ...292...90C}.
\item Mean $\rm M_{V,15} = -6.05\pm0.44$ magnitudes by \cite{2000AJ....120.2007D}.
\item $\rm M_{V,17} = -6.06\pm0.23$ magnitudes and $\rm M_{V,20} = -6.11\pm0.34$
for novae in M87 by \citet{2017arXiv170206988S}. 
\end{enumerate}

These show a larger scatter
than $\rm M_{V,12}=-6.616\pm0.043$ magnitudes and $\rm M_{V,15} = -6.22\pm0.07$ magnitudes 
for novae with $\rm 2.4 \le t_2 \le 86$ days that we estimate using the MMRD $\rm M_{V,0}$.

\section{Conclusions}
In this paper, the maximum (absolute) magnitude relation with decline time (MMRD) 
for novae has been revisited
in terms of its physical origin and calibration using observed parameters.     
The main points of our study can be summarised to be: 
\begin{itemize}
\item We derive the life-expansion relation and the MMRD.  We start 
with a rapid one-time injection of energy to the nova system which is
manifested in form of kinetic energy imparted
to the outer layers of the white dwarf.  This energy input causes the outer layers 
to be set in bulk outward motion as the ejecta and also ionized so that the ejecta starts
radiating.  The assumptions in this derivation are: (1) instantaneous injection of energy
(2) $\rm m_{ej}$ is same for all novae (3) $\rm E_{rad}$ is same for all novae.

\item We calibrate the MMRD using a two-step procedure wherein the first
step exclusively uses directly observed quantities of emission line widths 
(proxy for expansion velocity) and time the nova takes to decline by two magnitudes ($\rm t_2$)
from maximum.  This step quantifies the rate of change in peak absolute magnitude
($\rm M_{V,0}$) with $\rm t_2$.  In the second step, the zero-point of the MMRD is 
fine-tuned using carefully derived values of $\rm M_{V,0}$ and $\rm t_2$ 
for novae from literature.  The fully calibrated MMRD is
$\rm M_{V,0} = 2.16(\pm 0.15) log_{10} t_2 - 10.804(\pm 0.117)$ which has the lowest
uncertainty of all the relations which exist in literature. 

\item The new MMRD calibration is used to estimate $\rm M_{V,0}$ of the novae in our Galaxy, in M31
and in M87 and most novae are found to lie between $-6$ and $-10$ magnitudes.  
The novae in M31 give a distance modulus of $24.8$ magnitudes and those in
M87 give a distance modulus of $31.3$ magnitudes for the parent galaxy. 
The range of distances estimated for Galactic novae using the 
MMRD range from very closeby to $\sim 100$ kpc, the latter are unrealistic and hence wrong.

\item We investigate the reasons for such large distances to several
Galactic novae being estimated on using the MMRD $\rm M_{V,0}$ and the measured peak apparent
magnitudes $\rm m_{V,0}$.  We show that the major and frequent cause of error is
an incorrect estimate of $\rm m_{V,0}$ due to several novae being detected when
their brightness is already on the decline.  
The MMRD is not responsible for the discrepant distances.  Our work lends
strong support to the validity of the MMRD and the importance of an accurate 
calibration. 

\item The near-constancy of the absolute magnitude of several novae on a particular
day after outburst maximum is revisited.  Our study concludes that the scatter on the absolute
magnitude of novae with $\rm 2.4 \le t_2 \le 86$ days is the least on day 12 after maximum 
and this is $-6.616 \pm 0.043$ magnitudes.  Our MMRD was used to estimate $\rm M_{V,0}$
from which $\rm M_{V,12}$ was found.  This work gives independent
evidence to the MMRD presented in the paper being well-calibrated and 
a reliable estimator of the peak absolute magnitudes of novae.  

\item We, thus, conclude that the MMRD is indeed a very powerful method for
estimating the peak luminosities of nova outbursts and hence distances to novae. 
However the MMRD should be used to determine the distance to a nova only if
an accurate $\rm m_{V,0}$ is recorded which will generally be true for novae detected
before they start their decline.  The distances to other novae can be determined
from $\rm M_{V,12}$ and $\rm m_{V,12}$ if an accurate value of the latter is available. 

\item It might be possible to use the MMRD, with an independent calibration,
to determine the peak luminosities of other classes of transients
if they are suspected to share the same physics as the nova outburst especially
the instantaneous energy input which throws out an ejecta which starts shining. 

\item We end by suggesting a few ways in which we can improve the utility
of the MMRD and also gradually minimize the observational uncertainties so that 
we might then be able to examine the extent of the phase space in which the MMRD remains valid: 

(1) By uniformising the data on $\rm t_2$ for all novae.  This can be possible if
we can converge on the best methodologies for estimating $\rm t_2$.  Since a 
coarsely determined $\rm t_2$ has been sufficient for the MMRD,
one of the possible methods would be to fit the
observed visual light curve upto the point that it has declined by 6 magnitudes or
more with a polynomial and find $\rm t_2$ from the fit.  This would then
smooth over the small wiggles that are sometimes seen superposed on the smooth brightness
decline.  If such software is made available, for example, on the AAVSO website so
that users can use it to estimate $\rm t_2$ then it can lead to uniform results
for a given nova for all users.  Alternatively, AAVSO can fit 
the data on all novae in its database and make $\rm t_2$ available to all users.  

(2) A similar argument applies to determining $\rm m_{V,0}$.  
Since novae show a variety in the shape of the light curves near maximum, it is important to 
derive a robust algorithm for determining $\rm m_{V,0}$ which can then be
universally accepted and can lead to reducing the observational uncertainty on
these parameters.  We note that unlike $\rm t_2$, the peak luminosity that a nova outburst
achieves will be the largest reliable value of $\rm m_{V,0}$ that we record so that
it might not require any fitting of the light curve.
However it might require uniformising of the data that might
exist in different research papers or websites observed in different wavebands
and making the same data on a nova available at all locations for universal use.  
It is important to present actual data and avoid making any extrapolations on the data. 

(3) While it is useful to collect data on
a large number of novae, it should not lead to compromising the quality of the data.
In fact, high fidelity spectrophotometric data on fewer novae is preferable to lower fidelity
data on several novae. 

\item The MMRD, suggested by \citet{1936PASP...48..191Z}, 
correctly calibrated and observationally supported by 
\citet{1940ApJ....91..369M} and other
scientists remains valid.  Novae of several speed classes 
are also found to evolve to a near-constant luminosity
several days after outburst maximum due to their different rates of
decline as was suggested by \citet{1955Obs....75..170B}. 
These remain important insights into the nova phenomena.

\end{itemize}

\section*{Acknowledgements}
I gratefully acknowledge use of ADS abstracts, arXiv e-prints, AAVSO data, gnuplot, LaTeX,
Wikipedia and Google search engines enabled by the internet in this research. 

\bibliography{novae2016}

\section*{Appendix}
In the Appendix, we present tables in which the peak absolute magnitudes of novae in outburst
determined from the MMRD that
we obtain in the paper with the decline time $\rm t_2$ taken from literature are listed. 
For the extragalactic novae in the M31 and Virgo cluster, the distance modulus 
are estimated and listed.  For ease of referral, each table corresponds to the 
set of novae for which the $\rm t_2$ and $\rm m_{V,0}$
are taken from a particular reference or couple of references.  

\begin{table*}
\centering
\small
\caption{\small Using the MMRD relation calibrated in the paper to estimate the peak absolute magnitude
and then distance to novae.  
Data on $\rm t_2$, $\rm \Delta v$ and $\rm m_{V,0}$ are taken from \citet{2011ApJS..197...31S}.
These data were used in the step one of the calibration.   
The bracketted quantities in columns 6 and 7 indicate the $1\sigma$ error on the
absolute magnitude and distance to the nova.  All the novae have an outburst recorded 
between 1982 and 2010. }
\begin{tabular}{l|l|c|c|c|l|l}
\hline
&    &   \multicolumn{3}{c|}{\bf \citet{2011ApJS..197...31S}} & \multicolumn{2}{c}{\bf This MMRD} \\
{\bf No} & {\bf Nova} & {$\bf m_{V,0}$} &  {$\bf t_2$} & {$\bf \Delta v$} & {$\bf M_{V,0}$} & {\bf D} \\
&    &   mag  & days & kms$^{-1}$ & mag & kpc \\
\hline
\hline
1  &     CI Aql   & 8.83 &  32.0  & 2300 & $-7.6   (-7.9,    -7.2)$  &  18.9    (22.1,    16.1)   \\
3  &     GQ Mus   & 7.20 &  18.0  & 1000 & $-8.1   (-8.4,    -7.8)$  &  11.4    (13.2,    9.9 )  \\
4  &     IM Nor   & 7.84 &  50.0  & 1150 & $-7.1   (-7.5,    -6.8)$  &  9.9     (11.7,    8.3 )  \\
5  &     KT Eri   & 5.42 &  6.6   & 3000 & $-9.0   (-9.3,    -8.8)$  &  7.8     (8.7,     7.0)   \\
7 &     LMC 2000 & 11.45&  9.0    & 1700 & $-8.7   (-9.0,    -8.5)$  &  109.3   (123.2,   96.9)   \\
8 &     LMC 2005 & 11.50&  63.0   & 900 &  $-6.9   (-7.3,    -6.5)$  &  48.2    (57.7,    40.4)   \\
9 &     LMC 2009a& 10.60&  4.0    & 3900 & $ -9.5  (-9.7,    -9.3)$  &  104.9   (115.4,   95.3)   \\
12 &     RS Oph   & 4.50 &  7.9   & 3930 & $ -8.9  (-9.1,    -8.6)$  &  4.7     (5.3,     4.2)   \\
13 &     U Sco    & 8.05 &  1.2   & 7600 & $-10.6  (-10.8,   -10.5)$ &  54.5    (57.9,    51.4)*   \\
14 &     V1047 Cen& 8.50 &  6.0   & 840  & $ -9.1  (-9.4,    -8.9)$  &  33.5    (37.3,    30.1)*   \\
15 &     V1065 Cen& 8.20 &  11.0  & 2700 & $-8.6   (-8.8,    -8.3)$  &  22.4    (25.4,    19.8)*   \\
16 &     V1187 Sco& 7.40 &  7.0   & 3000 & $-9.0   (-9.2,    -8.7)$  &  18.9    (21.1,    16.9)   \\
17 &     V1188 Sco& 8.70 &  7.0   & 1730 & $-9.0   (-9.2,    -8.7)$  &  34.3    (38.4,    30.7)*   \\
18 &     V1213 Cen& 8.53 &  11.0  & 2300 & $-8.6   (-8.8,    -8.3)$  &  26.1    (29.6,    23.0)*   \\
19 &     V1280 Sco& 3.79 &  21.0  & 640  & $-7.9   (-8.3,    -7.6)$  &  2.2     (2.6,     1.9)   \\
20 &     V1281 Sco& 8.80 &  15.0  & 1800 & $-8.3   (-8.6,    -8.0)$  &  25.9    (29.6,    22.6)*   \\
21 &     V1309 Sco& 7.10 &  23.0  & 670  & $-7.9   (-8.2,    -7.5)$  &  9.8     (11.4,    8.5)   \\
22 &     V1494 Aql& 3.80 &  6.6   & 1200 & $-9.0   (-9.3,    -8.8)$  &  3.7     (4.1,     3.3)   \\
23 &     V1663 Aql& 10.50&  17.0  & 1900 & $-8.1   (-8.4,    -7.8)$  &  53.6    (61.6,    46.7)*   \\
24 &     V1974 Cyg& 4.30 &  17.0  & 2000 & $-8.1   (-8.4,    -7.8)$  &  3.1     (3.5,     2.7)   \\
25 &     V2361 Cyg& 9.30 &  6.0   & 3200 & $-9.1   (-9.4,    -8.9)$  &  48.4    (53.9,    43.4)*   \\
26 &     V2362 Cyg& 7.80 &  9.0   & 1850 & $-8.7   (-9.0,    -8.5)$  &  20.4    (22.9,    18.1)*   \\
27 &     V2467 Cyg& 6.70 &  7.0   & 950 & $-9.0    (-9.2,    -8.7)$  &  13.7    (15.3,    12.2)   \\
28 &     V2468 Cyg& 7.40 &  10.0  & 1000 & $-8.6   (-8.9,    -8.4)$  &  16.2    (18.3,    14.3)   \\
29 &     V2491 Cyg& 7.54 &  4.6   & 4860 & $-9.4   (-9.6,    -9.2)$  &  24.1    (26.7,    21.8)*   \\
30 &     V2487 Oph& 9.50 &  6.3   & 10000& $-9.1   (-9.3,    -8.8)$  &  51.9    (57.9,    46.6)*   \\
32 &     V2575 Oph& 11.10&  20.0  & 560 & $-8.0    (-8.3,    -7.7)$  &  65.9    (76.1,    57.1)*   \\
33 &     V2576 Oph& 9.20 &  8.0   & 1470 & $-8.9   (-9.1,    -8.6)$  &  40.8    (45.8,    36.3)*   \\
34 &     V2615 Oph& 8.52 &  26.5  & 800 & $-7.7    (-8.1,    -7.4)$  &  17.8    (20.7,    15.3)*   \\
35 &     V2670 Oph& 9.90 &  15.0  & 600 & $-8.3    (-8.6,    -8.0)$  &  42.9    (49.1,    37.5)*   \\
36 &     V2671 Oph& 11.10&  8.0   & 1210 & $-8.9   (-9.1,    -8.6)$  &  97.9    (109.9,   87.1)*   \\
37 &     V2672 Oph& 10.00&  2.3   & 8000 & $-10.0  (-10.2,   -9.9)$  &  101.0   (109.3,   93.4)*   \\
39 &     V382 Nor & 8.90 &  12.0  & 1850 & $-8.5   (-8.8,    -8.2)$  &  29.8    (33.9,    26.2)*   \\
40 &     V382 Vel & 2.85 &  4.5   & 2400 & $-9.4   (-9.6,    -9.2)$  &  2.8     (3.1,     2.5)   \\
41 &     V407 Cyg & 6.80 &  5.9   & 2760 & $-9.1   (-9.4,    -8.9)$  &  15.4    (17.2,    13.8)   \\
42 &     V458 Vul & 8.24 &  7.0   & 1750 & $-9.0   (-9.2,    -8.7)$  &  27.8    (31.1,    24.8)*   \\
43 &     V459 Vul & 7.57 &  18.0  & 910 & $-8.1    (-8.4,    -7.8)$  &  13.6    (15.6,    11.8)   \\
44 &     V4633 Sgr& 7.80 &  19.0  & 1700 & $-8.0   (-8.4,    -7.7)$  &  14.7    (17.0,    12.8)   \\
45 &     V4643 Sgr& 8.07 &  4.8   & 4700 & $-9.3   (-9.6,    -9.1)$  &  30.2    (33.4,    27.3)*   \\
46 &     V4743 Sgr& 5.00 &  9.0   & 2400 & $-8.7   (-9.0,    -8.5)$  &  5.6     (6.3,     5.0)   \\
47 &     V4745 Sgr& 7.41 &  8.6   & 1600 & $-8.8   (-9.0,    -8.5)$  &  17.3    (19.5,    15.4)   \\
49 &     V477S ct & 9.80 &  3.0   & 2900 & $-9.8   (-10.0,   -9.6)$  &  82.2    (89.6,    75.3)*   \\
50 &     V5114 Sgr& 8.38 &  11.0  & 2000 & $-8.6   (-8.8,    -8.3)$  &  24.4    (27.6,    21.5)   \\
51 &     V5115 Sgr& 7.70 &  7.0   & 1300 & $-9.0   (-9.2,    -8.7)$  &  21.7    (24.2,    19.4)   \\
52 &     V5116 Sgr& 8.15 &  6.5   & 970  & $-9.0   (-9.3,    -8.8)$  &  27.5    (30.7,    24.7)*   \\
53 &     V5558 Sgr& 6.53 &  125.0 & 1000 & $-6.3   (-6.7,    -5.8)$  &  3.6     (4.4,     3.0)   \\
54 &     V5579 Sgr& 5.56 &  7.0   & 1500 & $-9.0   (-9.2,    -8.7)$  &  8.1     (9.0,     7.2)   \\
55 &     V5583 Sgr& 7.43 &  5.0   & 2300 & $-9.3   (-9.5,    -9.1)$  &  22.1    (24.5,    20.0)*   \\
56 &     V574 Pup & 6.93 &  13.0  & 2800 & $-8.4   (-8.7,    -8.1)$  &  11.6    (13.3,    10.2)   \\
57 &     V597 Pup & 7.00 &  3.0   & 1800 & $-9.8   (-10.0,   -9.6)$  &  22.6    (24.7,    20.7)*   \\
60 &     V723 Cas & 7.10 &  263.0 & 600 &  $-5.6   (-6.1,    -5.1)$  &  3.4     (4.3,     2.8)   \\
\hline
\end{tabular}

* - marks the Galactic nova for which a distance $> 20$ kpc has been estimated.

\label{tab2}
\end{table*}

\begin{table*}[t]
\centering
\small
\caption{ Using the MMRD relation calibrated in the paper to estimate the peak absolute magnitude 
and then distance to novae.  
Data on $\rm t_2$ and $\rm m_{V,0}$ are taken from \citet{1960stat.conf..585M}. 
The bracketted quantities in columns 6 and 7 indicate the $1\sigma$ error on the
absolute magnitude and distance to the nova. }
\begin{tabular}{l|l|c|c|c|l|l}
\hline
&    &   & \multicolumn{2}{c|}{\bf \citet{1960stat.conf..585M}} & \multicolumn{2}{c}{\bf This MMRD} \\
{\bf No} & {\bf Nova} & {\bf Outburst} & {$\bf m_{V,0}$} &  {$\bf t_2$} & {$\bf M_{V,0}$} & {\bf D} \\
&    & year  & mag  & days & mag & kpc \\
\hline
\hline
1    &   V606 Aql & 1899 & 6.70  & 25.0   & -7.8    (-8.1,    -7.5)  &  7.9     (9.2,     6.8)   \\
2    &   V604 Aql & 1905 & 8.20  & 9.0    & -8.7    (-9.0,    -8.5)  &  24.5    (27.6,    21.7)*   \\
3    &   V603 Aql & 1918 & -1.10 & 4.0    & -9.5    (-9.7,    -9.3)  &  0.5     (0.5,     0.4)   \\
4    &   DO Aql   & 1925 & 8.60  & 300.0  & -5.5    (-5.9,    -5.0)  &  6.5     (8.1,     5.2)   \\
5    &   V356 Aql & 1936 & 7.10  & 145.0  & -6.1    (-6.6,    -5.7)  &  4.4     (5.4,     3.6)   \\
6    &   V368 Aql & 1936 & 5.00  & 5.0    & -9.3    (-9.5,    -9.1)  &  7.2     (8.0,     6.5)   \\
7    &   V528 Aql & 1945 & 7.00  & 17.0   & -8.1    (-8.4,    -7.8)  &  10.7    (12.3,    9.3)   \\
8    &   T Aur    & 1891 & 4.20  & 80.0   & -6.7    (-7.1,    -6.3)  &  1.5     (1.8,     1.3)   \\
9    &   T CrB    & 1866 & 2.00  & 3.0    & -9.8    (-10.0,   -9.6)  &  2.3     (2.5,     2.1)   \\
10   &   T CrB    & 1946 & 3.00  & 5.0    & -9.3    (-9.5,    -9.1)  &  2.9     (3.2,     2.6)   \\
11   &   Q Cyg    & 1876 & 3.00  & 5.0    & -9.3    (-9.5,    -9.1)  &  2.9     (3.2,     2.6)   \\
12   &   V476 Cyg & 1920 & 2.00  & 7.0    & -9.0    (-9.2,    -8.7)  &  1.6     (1.8,     1.4)   \\
13   &   V450 Cyg & 1942 & 7.90  & 91.0   & -6.6    (-7.0,    -6.2)  &  7.8     (9.5,     6.5)   \\
14   &   DM Gem   & 1903 & 5.00  & 6.0    & -9.1    (-9.4,    -8.9)  &  6.7     (7.4,     6.0)   \\
15   &   DN Gem   & 1912 & 3.50  & 16.0   & -8.2    (-8.5,    -7.9)  &  2.2     (2.5,     1.9)   \\
16   &   DQ Her   & 1934 & 1.40  & 67.0   & -6.9    (-7.3,    -6.5)  &  0.4     (0.5,     0.4)   \\
17   &   DI Lac   & 1910 & 4.60  & 20.0   & -8.0    (-8.3,    -7.7)  &  3.3     (3.8,     2.9)   \\
18   &   CP Lac   & 1936 & 2.10  & 5.0    & -9.3    (-9.5,    -9.1)  &  1.9     (2.1,     1.7)   \\
19   &   DK Lac   & 1950 & 5.40  & 19.0   & -8.0    (-8.4,    -7.7)  &  4.9     (5.6,     4.2)   \\
20   &   HR Lyr   & 1919 & 6.50  & 48.0   & -7.2    (-7.5,    -6.8)  &  5.4     (6.4,     4.6)   \\
21   &   V841 Oph & 1848 & 4.30  & 50.0   & -7.1    (-7.5,    -6.8)  &  1.9     (2.3,     1.6)   \\
22   &   RS Oph   & 1933 & 4.30  & 4.0    & -9.5    (-9.7,    -9.3)  &  5.8     (6.3,     5.2)   \\
23   &   V849 Oph & 1919 & 7.40  & 120.0  & -6.3    (-6.7,    -5.9)  &  5.5     (6.7,     4.5)   \\
24   &   GK Per   & 1901 & 0.20  & 6.0    & -9.1    (-9.4,    -8.9)  &  0.7     (0.8,     0.7)   \\
25   &   RR Pic   & 1925 & 1.20  & 80.0   & -6.7    (-7.1,    -6.3)  &  0.4     (0.5,     0.3)   \\
26   &   CP Pup   & 1942 & 0.40  & 5.0    & -9.3    (-9.5,    -9.1)  &  0.9     (1.0,     0.8)   \\
27   &   T Pyx    & 1944 & 7.00  & 100.0  & -6.5    (-6.9,    -6.1)  &  5.0     (6.0,     4.1)   \\
28   &   WZ Sge   & 1913 & 7.20  & 14.0   & -8.3    (-8.6,    -8.0)  &  12.8    (14.6,    11.2)   \\
29   &   WZ Sge   & 1946 & 7.80  & 18.0   & -8.1    (-8.4,    -7.8)  &  15.1    (17.4,    13.1)   \\
30   &   V1059 Sgr& 1898 & 4.90  & 10.0   & -8.6    (-8.9,    -8.4)  &  5.1     (5.8,     4.5)   \\
31   &   V999 Sgr & 1910 & 8.10  & 220.0  & -5.7    (-6.2,    -5.3)  &  5.9     (7.3,     4.7)   \\
32   &   V1017 Sgr & 1919& 7.00  & 70.0   & -6.8    (-7.2,    -6.4)  &  5.8     (7.0,     4.8)   \\
33   &   V630 Sgr & 1936 & 4.40  & 4.0    & -9.5    (-9.7,    -9.3)  &  6.0     (6.6,     5.5)   \\
34   &   EU Sct   & 1949 & 7.40  & 20.0   & -8.0    (-8.3,    -7.7)  &  12.0    (13.8,    10.4)   \\
35   &   RT Ser   & 1909 & 10.60 & 40.0   & -7.3    (-7.7,    -7.0)  &  38.8    (45.7,    32.9)*   \\
\hline
\end{tabular}

* - marks the Galactic nova for which a distance $> 20$ kpc has been estimated.

\label{tab3}
\end{table*}

\begin{table*}
\centering
\small
\caption{ Using the MMRD relation calibrated in the paper to estimate the peak absolute magnitude
and then distance to recurrent novae.
Data on $\rm t_2$ and $\rm m_{V,0}$ are taken from \citet{2010ApJS..187..275S}
The bracketted quantities in columns 7 and 8 indicate the $1\sigma$ error on the
absolute magnitude and distance to the nova. }
\begin{tabular}{l|l|c|c|c|c|l|l}
\hline
{\bf No} & {\bf Recurrent} & \multicolumn{4}{c|}{\bf  \citet{2010ApJS..187..275S}} & 
\multicolumn{2}{c}{\bf This MMRD} \\
& {\bf Nova} & {$\bf m_{V,0}$} &  {$\bf t_2$} & {$\bf M_{V,0}$} & {\bf D} &{$\bf M_{V,0}$} & {\bf D} \\ 
&    &   mag  & days &  mag & kpc &mag & kpc  \\
\hline
\hline
1    &   TPyx   &  6.40 &  32.0  & -7.1 & 3.2 &   -7.6    (-7.9,    -7.2) &   6.2     (7.2,     5.3)   \\
2    &   IMNor  &  8.50 &  50.0  & -7.0 & 3.4 &   -7.1    (-7.5,    -6.8) &   13.4    (15.9,    11.3)\\
3    &   CIAql  &  9.00 &  25.4  & -7.3 & 5.0 &   -7.8    (-8.1,    -7.4) &   22.6    (26.3,    19.4)*\\
4    &   V2487Oph& 9.50 &  6.2   & -9.6 & 32.4 &  -9.1    (-9.3,    -8.9) &   52.3    (58.3,    46.9)*\\
5    &   USco    & 7.50 &  1.2   & -10.7 & 37.7& -10.6   (-10.8,    -10.5)&   42.3    (44.9,    39.9)*\\
6    &   V394CrA & 7.20 &  2.4   & -10.2 & 24.4& -10.0   (-10.2,    -9.8) &   27.3    (29.6,    25.2)*\\
7    &   TCrB    & 2.50 &  4.0   & -9.9 & 3.2 &   -9.5    (-9.7,    -9.3) &   2.5     (2.8,     2.3)\\
8    &   RSOph   & 4.80 &  6.8   & -9.5 & 2.1 &   -9.0    (-9.2,    -8.8) &   5.8     (6.4,     5.2)\\
9    &   V745Sco & 9.40 &  6.2   & -9.6 & 14.1 &  -9.1    (-9.3,    -8.9) &   49.9    (55.7,    44.8)*\\
10   &   V3890Sgr& 8.10 &  6.4   & -9.6 & 7.6  &  -9.1    (-9.3,    -8.8) &   27.1    (30.2,    24.3)*\\
\hline
\end{tabular}

* - marks the Galactic nova for which a distance $> 20$ kpc has been estimated.

\label{tab4}
\end{table*}

\begin{table*}
\centering
\small
\caption{\small 
Using the MMRD relation calibrated in the paper to estimate the peak absolute magnitude
and then distance to novae.
Data on $\rm t_2$ and $\rm m_{V,0}$ are taken from \citet{2010AJ....140...34S}.
The bracketted quantities in columns 6 and 7 indicate the $1\sigma$ error on the
absolute magnitude and distance to the nova. }
\begin{tabular}{l|l|c|c|c|l|l}
\hline
&    &  &  \multicolumn{2}{c|}{\bf \citet{2010AJ....140...34S}} & \multicolumn{2}{c}{\bf This MMRD} \\
{\bf No} & {\bf Nova} & {\bf Outburst} & {$\bf m_{V,0}$} &  {$\bf t_2$} & {$\bf M_{V,0}$} & {\bf D} \\
&    & year &  mag  & days & mag & kpc  \\
\hline
\hline
1   &    OS And   & 1986 & 6.50  & 11.0  &  -8.6    (-8.8,    -8.3) &   10.3    (11.6,    9.0)    \\
2   &    CI Aql   & 2000 & 9.00  & 25.0  &  -7.8    (-8.1,    -7.5) &   22.7    (26.4,    19.6)*    \\
3   &    DO Aql   & 1925 & 8.50  & 295.0 &  -5.5    (-6.0,    -5.0) &   6.2     (7.8,     5.0)    \\
4   &    V356 Aql & 1936 & 7.00  & 127.0 &  -6.3    (-6.7,    -5.8) &   4.5     (5.5,     3.7)    \\
5   &    V528 Aql & 1945 & 6.90  &  16.0  &  -8.2    (-8.5,    -7.9) &   10.5    (12.0,    9.1)    \\
6   &    V603 Aql & 1918 & -0.50 &  5.0   &  -9.3    (-9.5,    -9.1) &   0.6     (0.6,     0.5)    \\
7   &    V1229 Aql& 1970 & 6.60 &  18.0  &  -8.1    (-8.4,    -7.8) &   8.7     (10.0,    7.5)    \\
8   &    V1370 Aql& 1982 & 7.70 &  15.0  &  -8.3    (-8.6,    -8.0) &   15.6    (17.8,    13.6)    \\
9   &    V1419 Aql& 1993 & 7.60 &  25.0  &  -7.8    (-8.1,    -7.5) &   11.9    (13.9,    10.3)    \\
10  &    V1425 Aql& 1995 & 8.00 &  27.0  &  -7.7    (-8.0,    -7.4) &   13.9    (16.2,    11.9)    \\
11  &    V1493 Aql& 1999 & 10.10&  9.0   &  -8.7    (-9.0,    -8.5) &   58.7    (66.2,    52.1)*    \\
12  &    V1494 Aql& 1999 & 4.10 &  8.0   &  -8.9    (-9.1,    -8.6) &   3.9     (4.4,     3.5)    \\
13  &    T Aur    & 1891 & 4.50 &  80.0  &  -6.7    (-7.1,    -6.3) &   1.7     (2.1,     1.4)    \\
14  &    V705 Cas & 1993 & 5.70 &  33.0  &  -7.5    (-7.9,    -7.2) &   4.4     (5.2,     3.8)    \\
15  &    V723 Cas & 1995 & 7.10 &  263.0 &  -5.6    (-6.1,    -5.1) &   3.4     (4.3,     2.8)    \\
16  &    V842 Cen & 1986 & 4.90 &  43.0  &  -7.3    (-7.6,    -6.9) &   2.7     (3.2,     2.3)    \\
17  &    V868 Cen & 1991 & 8.70 &  31.0  &  -7.6    (-7.9,    -7.2) &   18.1    (21.1,    15.4)    \\
18  &    V888 Cen & 1995 & 8.00 &  38.0  &  -7.4    (-7.7,    -7.0) &   12.0    (14.1,    10.2)    \\
19  &    V1039 Cen& 2001 & 9.30 &  25.0  &  -7.8    (-8.1,    -7.5) &   26.1    (30.4,    22.5)*    \\
20  &    BY Cir   & 1995 & 7.40 &  35.0  &  -7.5    (-7.8,    -7.1) &   9.4     (11.1,    8.0)    \\
21  &    DD Cir   & 1999 & 7.60 &  5.0   &  -9.3    (-9.5,    -9.1) &   23.9    (26.5,    21.6)*    \\
22  &    V693 CrA & 1981 & 7.00 &  10.0  &  -8.6    (-8.9,    -8.4) &   13.5    (15.2,    11.9)    \\
23  &    T CrB    & 1946 & 2.50 &  4.0   &  -9.5    (-9.7,    -9.3) &   2.5     (2.8,     2.3)    \\
24  &    V476 Cyg & 1920 & 1.90 &  6.0   &  -9.1    (-9.4,    -8.9) &   1.6     (1.8,     1.4)    \\
25  &    V1330 Cyg& 1970 & 9.90 &  161.0 &  -6.0    (-6.5,    -5.6) &   15.4    (18.9,    12.5)    \\
26  &    V1500 Cyg& 1975 & 1.90 &  2.0   &  -10.2   (-10.3,   -10.0)&   2.6     (2.8,     2.4)    \\
27  &    V1668 Cyg& 1978 & 6.20 &  11.0  &  -8.6    (-8.8,    -8.3) &   8.9     (10.1,    7.9)    \\
28  &    V1819 Cyg& 1986 & 9.30 &  95.0  &  -6.5    (-6.9,    -6.1) &   14.7    (17.7,    12.1)    \\
29  &    V1974 Cyg& 1992 & 4.30 &  19.0  &  -8.0    (-8.4,    -7.7) &   2.9     (3.4,     2.6)    \\
30  &    V2274 Cyg& 2001 & 11.50&  22.0  &  -7.9    (-8.2,    -7.6) &   76.0    (88.0,    65.6)*    \\
31  &    V2275 Cyg& 2001 & 6.90 &  3.0   &  -9.8    (-10.0,   -9.6) &   21.6    (23.6,    19.8)*    \\
32  &    V2362 Cyg& 2006 & 8.10 &  9.0   &  -8.7    (-9.0,    -8.5) &   23.4    (26.3,    20.7)*    \\
33  &    V2467 Cyg& 2007 & 7.40 &  8.0   &  -8.9    (-9.1,    -8.6) &   17.8    (20.0,    15.9)    \\
34  &    V2491 Cyg& 2008 & 7.50 &  4.0   &  -9.5    (-9.7,    -9.3) &   25.2    (27.7,    22.9)*    \\
35  &    HR Del   & 1967 & 3.60 &  167.0 &  -6.0    (-6.5,    -5.6) &   0.8     (1.0,     0.7)    \\
36  &    DN Gem   & 1912 & 3.60 &  16.0  &  -8.2    (-8.5,    -7.9) &   2.3     (2.6,     2.0)    \\
37  &    DQ Her   & 1934 & 1.60 &  76.0  &  -6.7    (-7.1,    -6.3) &   0.5     (0.6,     0.4)    \\
38  &    V446 Her & 1960 & 4.80 &  20.0  &  -8.0    (-8.3,    -7.7) &   3.6     (4.2,     3.1)    \\
39  &    V533 Her & 1963 & 3.00 &  30.0  &  -7.6    (-8.0,    -7.3) &   1.3     (1.6,     1.1)    \\
40  &    V827 Her & 1987 & 7.50 &  21.0  &  -7.9    (-8.3,    -7.6) &   12.3    (14.2,    10.6)    \\
41  &    V838 Her & 1991 & 5.30 &  1.0   &  -10.8   (-10.9,   -10.7)&   16.6    (17.5,    15.8)    \\
42  &    CP Lac   & 1936 & 2.00 &  5.0   &  -9.3    (-9.5,    -9.1) &   1.8     (2.0,     1.6)    \\
43  &    DK Lac   & 1950 & 5.90 &  55.0  &  -7.0    (-7.4,    -6.7) &   3.9     (4.6,     3.3)    \\
44  &    LZ Mus   & 1998 & 8.50 &  4.0   &  -9.5    (-9.7,    -9.3) &   39.9    (43.9,    36.2)*    \\
45  &    BT Mon   & 1939 & 8.10 &  118.0 &  -6.3    (-6.8,    -5.9) &   7.7     (9.4,     6.3)    \\
\hline
\end{tabular}
\label{tab5}
\end{table*}

\begin{table*}
\centering
\small
\begin{tabular}{l|l|c|c|c|l|l}
\hline
&    &  & \multicolumn{2}{c|}{\bf \citet{2010AJ....140...34S}} & \multicolumn{2}{c}{\bf This MMRD} \\
{\bf No} & {\bf Nova} & {\bf Outburst} & {$\bf m_{V,0}$} &  {$\bf t_2$} & {$\bf M_{V,0}$} & {\bf D} \\
&    & year & mag  & days & mag & kpc  \\
\hline
\hline
46  &    IM Nor   & 2002 & 8.50 &  50.0  &  -7.1    (-7.5,    -6.8) &   13.4    (15.9,    11.3)    \\
47  &    RS Oph   & 2006 & 4.80 &  7.0   &  -9.0    (-9.2,    -8.7) &   5.7     (6.4,     5.1)    \\
48  &    V849 Oph & 1919 & 7.60 &  140.0 &  -6.2    (-6.6,    -5.7) &   5.7     (6.9,     4.6)    \\
49  &    V2214 Oph& 1988 & 8.50 &  60.0  &  -7.0    (-7.3,    -6.6) &   12.4    (14.8,    10.4)    \\
50  &    V2264 Oph& 1991 & 10.00&  22.0  &  -7.9    (-8.2,    -7.6) &   38.1    (44.1,    32.9)*    \\
51  &    V2295 Oph& 1993 & 9.30 &  9.0   &  -8.7    (-9.0,    -8.5) &   40.6    (45.8,    36.0)*    \\
52  &    V2313 Oph& 1994 & 7.50 &  8.0   &  -8.9    (-9.1,    -8.6) &   18.6    (20.9,    16.6)    \\
53  &    V2487 Oph& 1998 & 9.50 &  6.0   &  -9.1    (-9.4,    -8.9) &   53.0    (59.1,    47.6)*    \\
54  &    V2540 Oph& 2002 & 8.10 &  66.0  &  -6.9    (-7.3,    -6.5) &   9.9     (11.8,    8.3)    \\
55  &    GK Per   & 1901 & 0.20 &  6.0   &  -9.1    (-9.4,    -8.9) &   0.7     (0.8,     0.7)    \\
56  &    RR Pic   & 1925 & 1.00 &  73.0  &  -6.8    (-7.2,    -6.4) &   0.4     (0.4,     0.3)    \\
57  &    CP Pup   & 1942 & 0.70 &  4.0   &  -9.5    (-9.7,    -9.3) &   1.1     (1.2,     1.0)    \\
58  &    V351 Pup & 1991 & 6.40 &  9.0   &  -8.7    (-9.0,    -8.5) &   10.7    (12.0,    9.5)    \\
59  &    V445 Pup & 2000 & 8.60 &  215.0 &  -5.8    (-6.2,    -5.3) &   7.5     (9.3,     6.0)    \\
60  &    V574 Pup & 2004 & 7.00 &  12.0  &  -8.5    (-8.8,    -8.2) &   12.4    (14.1,    10.9)    \\
61  &    T Pyx    & 1967 & 6.40 &  32.0  &  -7.6    (-7.9,    -7.2) &   6.2     (7.2,     5.3)    \\
62  &    HS Sge   & 1977 & 7.20 &  15.0  &  -8.3    (-8.6,    -8.0) &   12.4    (14.2,    10.8)    \\
63  &    V732 Sgr & 1936 & 6.40 &  65.0  &  -6.9    (-7.3,    -6.5) &   4.5     (5.4,     3.8)    \\
64  &    V3890 Sgr& 1990 & 8.10 &  6.0   &  -9.1    (-9.4,    -8.9) &   27.8    (31.0,    25.0)*    \\
65  &    V4021 Sgr& 1977 &8.90 &  56.0  &  -7.0    (-7.4,    -6.6) &   15.3    (18.3,    12.9)    \\
66  &    V4160 Sgr& 1991 & 7.00 &  2.0   &  -10.2   (-10.3,   -10.0)&   27.0    (29.1,    25.0)*    \\
67  &    V4169 Sgr& 1992 & 7.90 &  24.0  &  -7.8    (-8.1,    -7.5) &   13.9    (16.2,    12.0)    \\
68  &    V4444 Sgr& 1999 & 7.60 &  5.0   &  -9.3    (-9.5,    -9.1) &   23.9    (26.5,    21.6)    \\
69  &    V4633 Sgr& 1998 & 7.40 &  17.0  &  -8.1    (-8.4,    -7.8) &   12.9    (14.8,    11.2)    \\
70  &    V4643 Sgr& 2001 & 7.70 &  3.0   &  -9.8    (-10.0,   -9.6) &   31.2    (34.1,    28.6)*    \\
71  &    V4739 Sgr& 2001 & 7.20 &  2.0   &  -10.2   (-10.3,   -10.0)&   29.6    (31.9,    27.4)*    \\
72  &    V4740 Sgr& 2001 & 6.70 &  18.0  &  -8.1    (-8.4,    -7.8) &   9.1     (10.5,    7.9)    \\
73  &    V4742 Sgr& 2002 & 7.90 &  9.0   &  -8.7    (-9.0,    -8.5) &   21.3    (24.0,    18.9)*    \\
74  &    V4743 Sgr& 2002 & 5.00 &  6.0   &  -9.1    (-9.4,    -8.9) &   6.7     (7.4,     6.0)    \\
75  &    V4745 Sgr& 2003 & 7.30 &  79.0  &  -6.7    (-7.1,    -6.3) &   6.3     (7.6,     5.3)    \\
76  &    V5114 Sgr& 2004 & 8.10 &  9.0   &  -8.7    (-9.0,    -8.5) &   23.4    (26.3,    20.7)*    \\
77  &    V5115 Sgr& 2005 & 7.90 &  7.0   &  -9.0    (-9.2,    -8.7) &   23.8    (26.6,    21.2)*    \\
78  &    V5116 Sgr& 2005 & 7.60 &  12.0  &  -8.5    (-8.8,    -8.2) &   16.4    (18.6,    14.4)    \\
79  &    U Sco    & 1999 & 7.50 &  1.0   &  -10.8   (-10.9,   -10.7)&   45.8    (48.3,    43.4)*    \\
80  &    V992 Sco & 1992 & 7.70 &  100.0 &  -6.5    (-6.9,    -6.1) &   6.9     (8.3,     5.7)    \\
81  &    V1186 Sco& 2004 & 9.70 &  12.0  &  -8.5    (-8.8,    -8.2) &   43.1    (49.0,    37.9)*    \\
82  &    V1187 Sco& 2004 & 9.80 &  10.0  &  -8.6    (-8.9,    -8.4) &   48.8    (55.2,    43.2)*    \\
83  &    V1188 Sco& 2005 & 8.90 &  11.0  &  -8.6    (-8.8,    -8.3) &   31.0    (35.1,    27.3)*    \\
84  &    V373 Sct & 1975 & 6.10 &  47.0  &  -7.2    (-7.6,    -6.8) &   4.6     (5.4,     3.8)    \\
85  &    V443 Sct & 1989 & 8.50 &  33.0  &  -7.5    (-7.9,    -7.2) &   16.0    (18.8,    13.7)    \\
86  &    FH Ser   & 1970 & 4.50 &  49.0  &  -7.2    (-7.5,    -6.8) &   2.1     (2.5,     1.8)    \\
87  &    LW Ser   & 1978 & 8.30 &  32.0  &  -7.6    (-7.9,    -7.2) &   14.8    (17.3,    12.6)    \\
88  &    V382 Vel & 1999 & 2.80 &  6.0   &  -9.1    (-9.4,    -8.9) &   2.4     (2.7,     2.2)    \\
89  &    LV Vul   & 1968 & 4.50 &  20.0  &  -8.0    (-8.3,    -7.7) &   3.2     (3.6,     2.7)    \\
90  &    NQ Vul   & 1976 & 6.20 &  21.0  &  -7.9    (-8.3,    -7.6) &   6.8     (7.8,     5.8)    \\
91  &    PW Vul   & 1984 & 6.40 &  44.0  &  -7.3    (-7.6,    -6.9) &   5.4     (6.4,     4.6)    \\
92  &    QU Vul   & 1984 & 5.30 &  20.0  &  -8.0    (-8.3,    -7.7) &   4.6     (5.3,     3.9)    \\
93  &    QV Vul   & 1987 & 7.10 &  37.0  &  -7.4    (-7.8,    -7.1) &   8.0     (9.4,     6.8)    \\
\hline
\end{tabular}

* - marks the Galactic nova for which a distance $> 20$ kpc has been estimated.

\end{table*}

\begin{table*}[t]
\centering
\small
\caption{\small 
Using the MMRD relation calibrated in the paper to estimate the peak absolute magnitude
and then distance to novae in M31.
Data on $\rm t_2$ and $\rm m_{V,0}$ are taken from \citet{1956AJ.....61...15A,1964AnAp...27..498R}.  
The bracketted quantities in columns 4 and 5 indicate the $1\sigma$ error on the
absolute magnitude and distance to the nova. }
\begin{tabular}{l|c|c|l|l}
\hline
\hline
 \multicolumn{3}{c|}{\bf \citet{1956AJ.....61...15A}} & \multicolumn{2}{c}{\bf This MMRD}\\
{\bf No}  & {$\bf m_{pg,0}$} &  {$\bf t_2$} & {$\bf M_{0}$} & {$ \bf m_{0}-M_{0}$} \\
&       mag  & days & mag & mag \\
\hline
1     &   15.70 & 2.0  &   -10.2   (-10.3,   -10.0) &  25.9    (26.0,    25.7) \\
2     &   15.70 & 2.0  &   -10.2   (-10.3,   -10.0) &  25.9    (26.0,    25.7) \\
3     &   15.90 & 5.0  &   -9.3    (-9.5,    -9.1)  &  25.2    (25.4,    25.0) \\
4     &   18.20 & 10.0 &   -8.6    (-8.9,    -8.4)  &  26.8    (27.1,    26.6) \\
5     &   15.90 & 12.0 &   -8.5    (-8.8,    -8.2)  &  24.4    (24.7,    24.1) \\
6     &   16.00 & 9.0  &   -8.7    (-9.0,    -8.5)  &  24.7    (25.0,    24.5) \\
7     &   15.90 & 13.0 &   -8.4    (-8.7,    -8.1)  &  24.3    (24.6,    24.0) \\
8     &   16.00 & 9.0  &   -8.7    (-9.0,    -8.5)  &  24.7    (25.0,    24.5) \\
9     &   16.00 & 11.0 &   -8.6    (-8.8,    -8.3)  &  24.6    (24.8,    24.3) \\
10    &   16.00 & 7.0  &   -9.0    (-9.2,    -8.7)  &  25.0    (25.2,    24.7) \\
12    &   16.10 & 11.0 &   -8.6    (-8.8,    -8.3)  &  24.7    (24.9,    24.4) \\
13    &   17.00 & 26.0 &   -7.7    (-8.1,    -7.4)  &  24.7    (25.1,    24.4) \\
14    &   16.20 & 12.0 &   -8.5    (-8.8,    -8.2)  &  24.7    (25.0,    24.4) \\
15    &   16.40 & 16.0 &   -8.2    (-8.5,    -7.9)  &  24.6    (24.9,    24.3) \\
16    &   16.70 & 13.0 &   -8.4    (-8.7,    -8.1)  &  25.1    (25.4,    24.8) \\
17    &   17.20 & 29.0 &   -7.6    (-8.0,    -7.3)  &  24.8    (25.2,    24.5) \\
18    &   17.50 & 34.0 &   -7.5    (-7.8,    -7.1)  &  25.0    (25.3,    24.6) \\
19    &   17.60 & 29.0 &   -7.6    (-8.0,    -7.3)  &  25.2    (25.6,    24.9) \\
20    &   17.20 & 33.0 &   -7.5    (-7.9,    -7.2)  &  24.7    (25.1,    24.4) \\
21    &   17.40 & 27.0 &   -7.7    (-8.0,    -7.4)  &  25.1    (25.4,    24.8) \\
22    &   17.60 & 30.0 &   -7.6    (-8.0,    -7.3)  &  25.2    (25.6,    24.9) \\
23    &   17.40 & 43.0 &   -7.3    (-7.6,    -6.9)  &  24.7    (25.0,    24.3) \\
24    &   17.80 & 34.0 &   -7.5    (-7.8,    -7.1)  &  25.3    (25.6,    24.9) \\
25    &   17.60 & 33.0 &   -7.5    (-7.9,    -7.2)  &  25.1    (25.5,    24.8) \\
26    &   18.00 & 47.0 &   -7.2    (-7.6,    -6.8)  &  25.2    (25.6,    24.8) \\
28    &   17.80 & 105.0&   -6.4    (-6.9,    -6.0)  &  24.2    (24.7,    23.8) \\
29    &   18.00 & 118.0&   -6.3    (-6.8,    -5.9)  &  24.3    (24.8,    23.9) \\
30    &   18.10 & 118.0&   -6.3    (-6.8,    -5.9)  &  24.4    (24.9,    24.0) \\
\hline
 \multicolumn{3}{c|}{\bf \citet{1964AnAp...27..498R}} & &\\
\hline
4      &     16.80 & 15.4  &  -8.2    (-8.5,    -7.9)  &  25.0    (25.3,    24.7)   \\
5      &     15.50 & 6.9   &  -9.0    (-9.2,    -8.7)  &  24.5    (24.7,    24.2)   \\
6      &     16.60 & 13.9  &  -8.3    (-8.6,    -8.0)  &  24.9    (25.2,    24.6)   \\
7      &     16.90 & 5.0   &  -9.3    (-9.5,    -9.1)  &  26.2    (26.4,    26.0)   \\
9      &     17.10 & 13.3  &  -8.4    (-8.7,    -8.1)  &  25.5    (25.8,    25.2)   \\
12     &     17.60 & 40.0  &  -7.3    (-7.7,    -7.0)  &  24.9    (25.3,    24.6)   \\
13     &     16.90 & 16.7  &  -8.2    (-8.5,    -7.9)  &  25.1    (25.4,    24.8)   \\
14     &     17.70 & 25.0  &  -7.8    (-8.1,    -7.5)  &  25.5    (25.8,    25.2)   \\
15     &     16.30 & 20.0  &  -8.0    (-8.3,    -7.7)  &  24.3    (24.6,    24.0)   \\
16     &     16.40 & 9.1   &  -8.7    (-9.0,    -8.5)  &  25.1    (25.4,    24.9)   \\
17     &     17.60 & 11.8  &  -8.5    (-8.8,    -8.2)  &  26.1    (26.4,    25.8)   \\
18     &     17.10 & 22.2  &  -7.9    (-8.2,    -7.6)  &  25.0    (25.3,    24.7)   \\
19     &     17.90 & 80.0  &  -6.7    (-7.1,    -6.3)  &  24.6    (25.0,    24.2)   \\
20     &     16.90 & 22.2  &  -7.9    (-8.2,    -7.6)  &  24.8    (25.1,    24.5)   \\
21     &     16.70 & 22.2  &  -7.9    (-8.2,    -7.6)  &  24.6    (24.9,    24.3)   \\
23     &     17.40 & 13.3  &  -8.4    (-8.7,    -8.1)  &  25.8    (26.1,    25.5)   \\
24     &     16.10 & 10.0  &  -8.6    (-8.9,    -8.4)  &  24.7    (25.0,    24.5)   \\
27     &     17.00 & 28.6  &  -7.7    (-8.0,    -7.3)  &  24.7    (25.0,    24.3)   \\
28     &     14.85 & 8.0   &  -8.9    (-9.1,    -8.6)  &  23.7    (24.0,    23.5)   \\
29     &     16.90 & 6.7   &  -9.0    (-9.3,    -8.8)  &  25.9    (26.2,    25.7)   \\
30     &     16.20 & 11.8  &  -8.5    (-8.8,    -8.2)  &  24.7    (25.0,    24.4)   \\
33     &     16.60 & 5.0   &  -9.3    (-9.5,    -9.1)  &  25.9    (26.1,    25.7)   \\
34     &     17.20 & 16.7  &  -8.2    (-8.5,    -7.9)  &  25.4    (25.7,    25.1)   \\
36     &     17.00 & 8.0   &  -8.9    (-9.1,    -8.6)  &  25.9    (26.1,    25.6)   \\
37     &     17.10 & 18.2  &  -8.1    (-8.4,    -7.8)  &  25.2    (25.5,    24.9)   \\
38     &     16.90 & 14.3  &  -8.3    (-8.6,    -8.0)  &  25.2    (25.5,    24.9)   \\
41     &     17.90 & 100.0 &  -6.5    (-6.9,    -6.1)  &  24.4    (24.8,    24.0)   \\
42     &     16.95 & 28.6  &  -7.7    (-8.0,    -7.3)  &  24.6    (24.9,    24.3)   \\
46     &     16.70 & 16.0  &  -8.2    (-8.5,    -7.9)  &  24.9    (25.2,    24.6)   \\
\hline
\end{tabular}
\label{tab6}
\end{table*}

\begin{table*}
\centering
\small
\caption{\small Using the MMRD relation calibrated in the paper to estimate the peak absolute magnitude
and then distance to novae in Virgo cluster galaxies. 
Data on $\rm t_2$ and $\rm m_{V,0}$ are taken from \citet{1987ApJ...318..507P,2016ApJS..227....1S}.
The bracketted quantities in columns 4 and 5 indicate the $1\sigma$ error on the
absolute magnitude and distance to the nova.}
\begin{tabular}{l|c|c|l|l}
\hline
\hline
 \multicolumn{3}{c|}{\bf Pritchet \& van den Bergh} & \multicolumn{2}{c}{This MMRD}\\
 \multicolumn{3}{c|}{\bf (1987)} & &\\
{\bf Nova}  & {$\bf m_{0}$} &  {$\bf t_2$} & {$\bf M_{0}$} & {$ \bf m_{0}-M_{0}$} \\
&       mag  & days & mag & mag \\
\hline
    NGC4365a   &      24.36 & 18.0 &   -8.1    (-8.4,    -7.8) &   32.5    (32.8,    32.1)    \\
    NGC4472Wa  &      22.74 & 20.0 &   -8.0    (-8.3,    -7.7) &   30.7    (31.0,    30.4)    \\
    NGC4472Wb  &      21.75 & 6.0  &   -9.1    (-9.4,    -8.9) &   30.9    (31.1,    30.6)    \\
    NGC4472Wc  &      24.44 & 12.5 &   -8.4    (-8.7,    -8.2) &   32.9    (33.2,    32.6)    \\
    NGC4472We  &      24.04 & 54.0 &   -7.1    (-7.4,    -6.7) &   31.1    (31.5,    30.7)    \\
    NGC4472Eb  &      23.21 & 11.0 &   -8.6    (-8.8,    -8.3) &   31.8    (32.0,    31.5)    \\
\hline
 \multicolumn{3}{c|}{\bf Novae in M87}  & &\\
 \multicolumn{3}{c|}{\bf  \citet{2016ApJS..227....1S}} & &\\
\hline
1     &    21.84 & 15.2 &   -8.2    (-8.5,    -8.0)  &  30.1    (30.4,    29.8) \\
2     &    21.85 & 11.2 &   -8.5    (-8.8,    -8.3)  &  30.4    (30.7,    30.1) \\
3     &    22.71 & 2.0  &   -10.1   (-10.3,   -10.0) &  32.9    (33.0,    32.7) \\
4     &    22.28 & 7.7  &   -8.9    (-9.1,    -8.6)  &  31.2    (31.4,    30.9) \\
5     &    22.74 & 17.1 &   -8.1    (-8.4,    -7.8)  &  30.9    (31.2,    30.6) \\
6     &    22.70 & 11.2 &   -8.5    (-8.8,    -8.3)  &  31.2    (31.5,    31.0) \\
7     &    23.61 & 9.3  &   -8.7    (-9.0,    -8.4)  &  32.3    (32.6,    32.1) \\
8     &    22.95 & 3.7  &   -9.6    (-9.8,    -9.4)  &  32.5    (32.7,    32.3) \\
10    &    23.51 & 22.3 &   -7.9    (-8.2,    -7.6)  &  31.4    (31.7,    31.1) \\
11    &    23.57 & 28.9 &   -7.6    (-8.0,    -7.3)  &  31.2    (31.6,    30.9) \\
12    &    23.58 & 33.1 &   -7.5    (-7.9,    -7.2)  &  31.1    (31.4,    30.8) \\
13    &    23.67 & 6.7  &   -9.0    (-9.3,    -8.8)  &  32.7    (32.9,    32.5) \\
14    &    23.74 & 31.5 &   -7.6    (-7.9,    -7.2)  &  31.3    (31.6,    31.0) \\
16    &    23.77 & 32.6 &   -7.5    (-7.9,    -7.2)  &  31.3    (31.6,    31.0) \\
17    &    23.81 & 30.4 &   -7.6    (-7.9,    -7.3)  &  31.4    (31.8,    31.1) \\
18    &    23.83 & 3.8  &   -9.6    (-9.8,    -9.4)  &  33.4    (33.6,    33.2) \\
19    &    23.90 & 9.0  &   -8.7    (-9.0,    -8.5)  &  32.6    (32.9,    32.4) \\
20    &    23.94 & 36.3 &   -7.4    (-7.8,    -7.1)  &  31.4    (31.7,    31.0) \\
21    &    24.07 & 29.9 &   -7.6    (-8.0,    -7.3)  &  31.7    (32.0,    31.3) \\
22    &    24.14 & 8.3  &   -8.8    (-9.1,    -8.6)  &  33.0    (33.2,    32.7) \\
23    &    24.16 & 8.0  &   -8.9    (-9.1,    -8.6)  &  33.0    (33.3,    32.8) \\
\hline
\end{tabular}
\label{tab7}
\end{table*}

\end{document}